\documentclass[12pt]{article}
\pdfoutput=1
\usepackage{amsmath}
\usepackage{amssymb}
\usepackage{graphicx}
\usepackage{subcaption}
\usepackage{url}
\usepackage{comment}

\usepackage[sort&compress, numbers, merge]{natbib}
\usepackage{enumerate}
\usepackage{multirow}

\usepackage[colorlinks=true, linkcolor=blue, citecolor=blue,
urlcolor=black]{hyperref}

\begin{document}

\begin{titlepage}
\begin{center}

\hfill DESY 16-005 \\
\hfill APCTP Pre2016-003\\
\hfill UT-16-02\\
\hfill \today

\vspace{1.0cm}
{\large\bf Testing ATLAS $Z$+MET Excess with LHC Run 2}

\vspace{1.0cm}
{\bf Xiaochuan Lu}$^{(a)}$,
{\bf Satoshi Shirai}$^{(b)}$ and
{\bf Takahiro Terada}$^{(c,d)}$

\vspace{1.0cm}
{\it
$^{(a)}${Department of Physics, University of California, Davis, California 95616, USA} \\
$^{(b)}${Deutsches Elektronen-Synchrotron (DESY), 22607 Hamburg, Germany} \\
$^{(c)}${Department of Physics, University of Tokyo, Tokyo 113-0033, Japan\\
$^{(d)}${Asia Pacific Center for Theoretical Physics (APCTP), Pohang, 37673, Korea}}
}

\vspace{1cm}
\abstract{
The ATLAS collaboration reported a 3$\sigma$ excess in the search of events containing on-$Z$ dilepton, jets, and large missing momentum (MET) in the 8~TeV LHC run. Motivated by this excess, many models of new physics have been proposed. Recently, the ATLAS and CMS collaborations reported new results for similar $Z$+MET channels in the 13~TeV run. In this paper, we comprehensively discuss the consistency between the proposed models and the LHC results of Run~1 and Run~2. We find that in models with heavy gluino production, there is generically some tension between the 8 TeV and 13 TeV results. On the other hand, models with light squark production provide relatively better fitting to both results.
}
\end{center}
\end{titlepage}

\setcounter{footnote}{0}

\section{Introduction}
The ATLAS collaboration reported an excess in the Supersymmetry (SUSY) search for events containing on-$Z$ dilepton with large missing transverse momentum (MET) and jet activity at the 8 TeV run~\cite{Aad:2015wqa}. In this search, 16 (13) events were observed in the di-electron (di-muon) signal channel, while the number of the expected Standard Model (SM) background is  $4.2\pm1.6$ ($6.4\pm 2.2$). This reads a 3.0$\sigma$ deviation from the SM background. Assuming that the ATLAS $Z$+MET excess comes from new physics, the observed \textit{visible cross section}, which is the product of the cross section $\sigma$ and the acceptance rate $\epsilon$ of the signal channel, is estimated as
\begin{align}
(\epsilon\sigma)_{8\text{TeV}}^\text{NP,obs} = 0.9 \pm 0.3~\text{fb}. \label{eqn:visiblexsec8TeV}
\end{align}

Although this excess has yet to be statistically significant to be an evidence for new physics, it gathered much attention. Many models of new physics have been proposed to explain this excess~\cite{Barenboim:2015afa, *Mitsou:2015vbx, Allanach:2015xga, Ellwanger:2015hva, Cao:2015ara, Harigaya:2015pma, Lu:2015wwa, Liew:2015hsa, Kobakhidze:2015dra, Cao:2015zya, Cahill-Rowley:2015cha, Collins:2015boa, Vignaroli:2015ama}. Most of these models are based on SUSY, but some non-SUSY models are also proposed. Generically, these models also predict other signals, because the $Z$ boson has a hadronic decay branching fraction much larger than its dilepton decay. For instance, the jets+MET+zero-lepton search is expected to place stringent constraints on them. However, many proposed models can circumvent these constraints. Typically, they have  compressed mass spectra and thus, reduce the MET and jet activities. This feature is also consistent with the detailed analysis of the ATLAS $Z$+MET excess --- the ATLAS collaboration provides more detailed information on the $Z$+MET events, including the distributions of MET, $H_\text{T}$ (a scalar sum of the transverse momenta of the signal jets and leptons) and jet multiplicity~\cite{Aad:2015wqa}. These distributions show that  lower jet activity is favoured, which is consistent with the negative results of jets+MET+zero-lepton search.

Another subtlety is the $Z$+MET search by the CMS at Run 1~\cite{Khachatryan:2015lwa}. The CMS collaboration has an analogous search for events with on-$Z$ dilepton, jets, and MET. While these channels resemble the ATLAS channels, no excess is found. This complicates the situation and dictates further tests on the ATLAS $Z$+MET excess. However, as the event selections in the ATLAS and CMS searches at Run 1 are not exactly the same --- for instance, no $H_{\text{T}}$ cut is required in the CMS search --- some models can survive the CMS constraints.

Recently, the ATLAS and CMS collaborations reported the first Run 2 results of the searches for the $Z$+MET events~\cite{ATLAS13TeV, CMS:2015bsf}, using data taken in 2015. Although these searches are performed with slightly severer lepton selection criteria, they are essentially very similar to the ATLAS $Z$+MET search at Run 1. Interestingly, the ATLAS collaboration again reports an excess, but the CMS result is still consistent with the SM background. In the ATLAS Run 2 $Z$+MET search, 21 events are observed in the dilepton channel, while $10.3\pm2.3$ SM backgrounds are expected. This amounts to a 2.2$\sigma$ excess. In Table~\ref{tab:observation}, we summarize the observations by the ATLAS and CMS collaborations.

\begin{table}[t]
  \begin{center}
    \caption{The number of observed events and expected SM background in the $Z$+MET searches.
    The distribution of the number of $b$-jets among the leading three jets is taken from
     Figure 5.25 of Ref.~\cite{Schreyer:2055513}.
    }
    \begin{tabular}{|c|c|c|c|c|} \hline
       & $n_{b\text{-jets}}^{}$ & SM background & Observed & Reference\\ \hline
      ATLAS 8 TeV (20.3 fb{$^{-1}$})& & $10.6\pm 3.2$  & 29 & \cite{Aad:2015wqa} \\
      & 0   & $(7.5\pm1.4)$   & 18  &\cite{Schreyer:2055513}\\
      & 1   & $(4.7\pm0.5)$   &  8  & \\
      & 2   & $(1.5\pm0.9)$   &  3  &\\
      & 3   &  $(0)$  &  0  &\\\hline
            ATLAS 13 TeV (3.2 fb{$^{-1}$})& & $10.3\pm 2.3$  & 21 & \cite{ATLAS13TeV} \\ \hline
            CMS 13 TeV (ATLAS-like) (2.2 fb{$^{-1}$})& & $12.0^{+4.0}_{-2.8}$  & 12 & \cite{CMS:2015bsf}  \\ \hline

    \end{tabular}
    \label{tab:observation}
  \end{center}
\end{table}

In this paper, we study the consistency between the models proposed to explain the 8 TeV ATLAS $Z$+MET excess and the 13 TeV results of the ATLAS and CMS collaborations. This comparison provides robust tests of such models --- for certain new physics models, accompanying signals such as jets+MET+zero-lepton events, or some other optimized events cut, might be more constraining. However, this kind of indirect constraints may be circumvented by tuning some parameters, and hence less robust than the direct comparison of the same channels at 8/13 TeV.

\begin{figure}[h]
\centering
 \subcaptionbox{Cross section ratios 13 TeV/ 8 TeV. \label{subfig:cross_ratio}}{\includegraphics[width=0.47\textwidth]{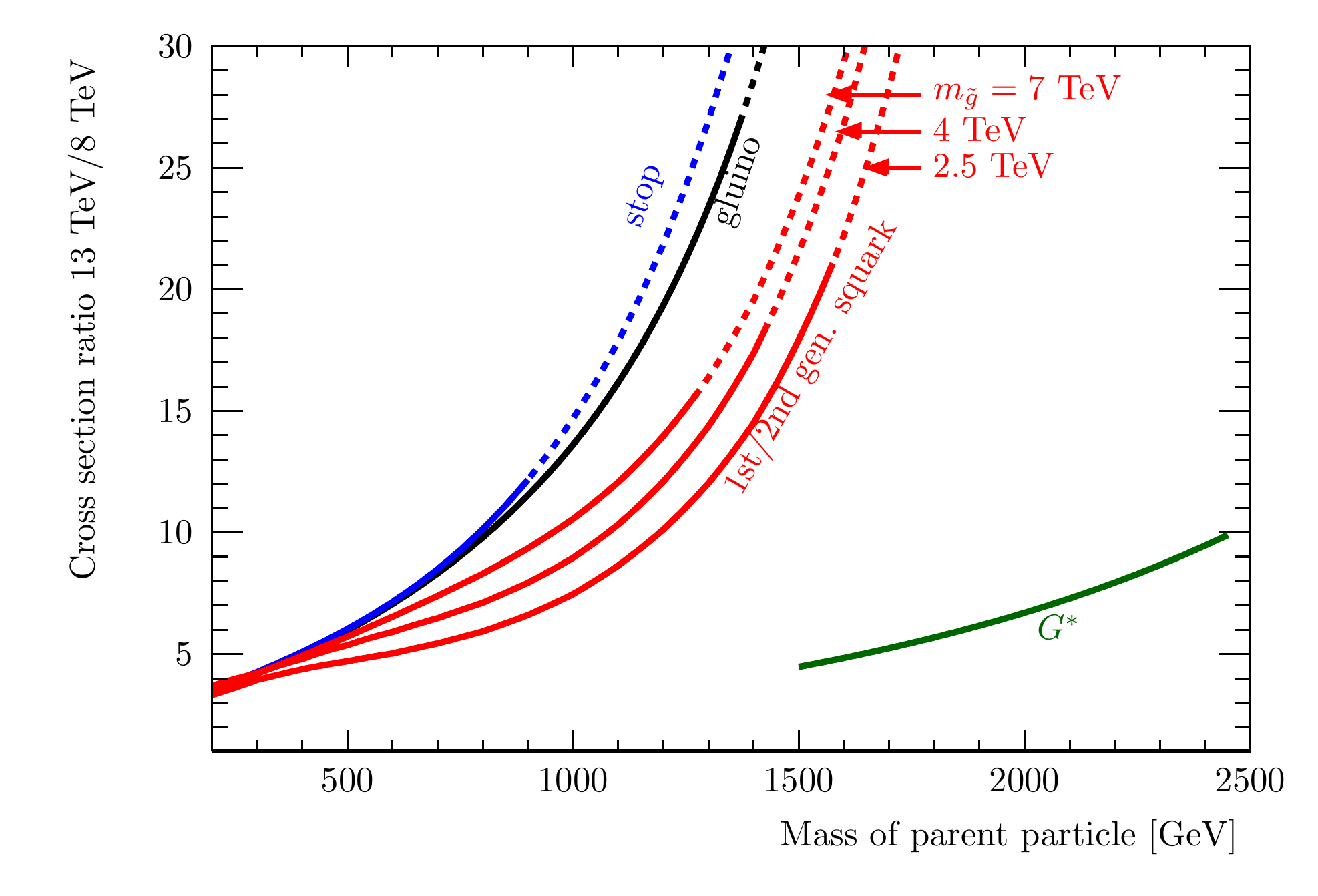}}
 \subcaptionbox{Acceptance ratios 13 TeV/ 8 TeV.
 \label{subfig:acc_ratio}}{\includegraphics[width=0.47\textwidth]{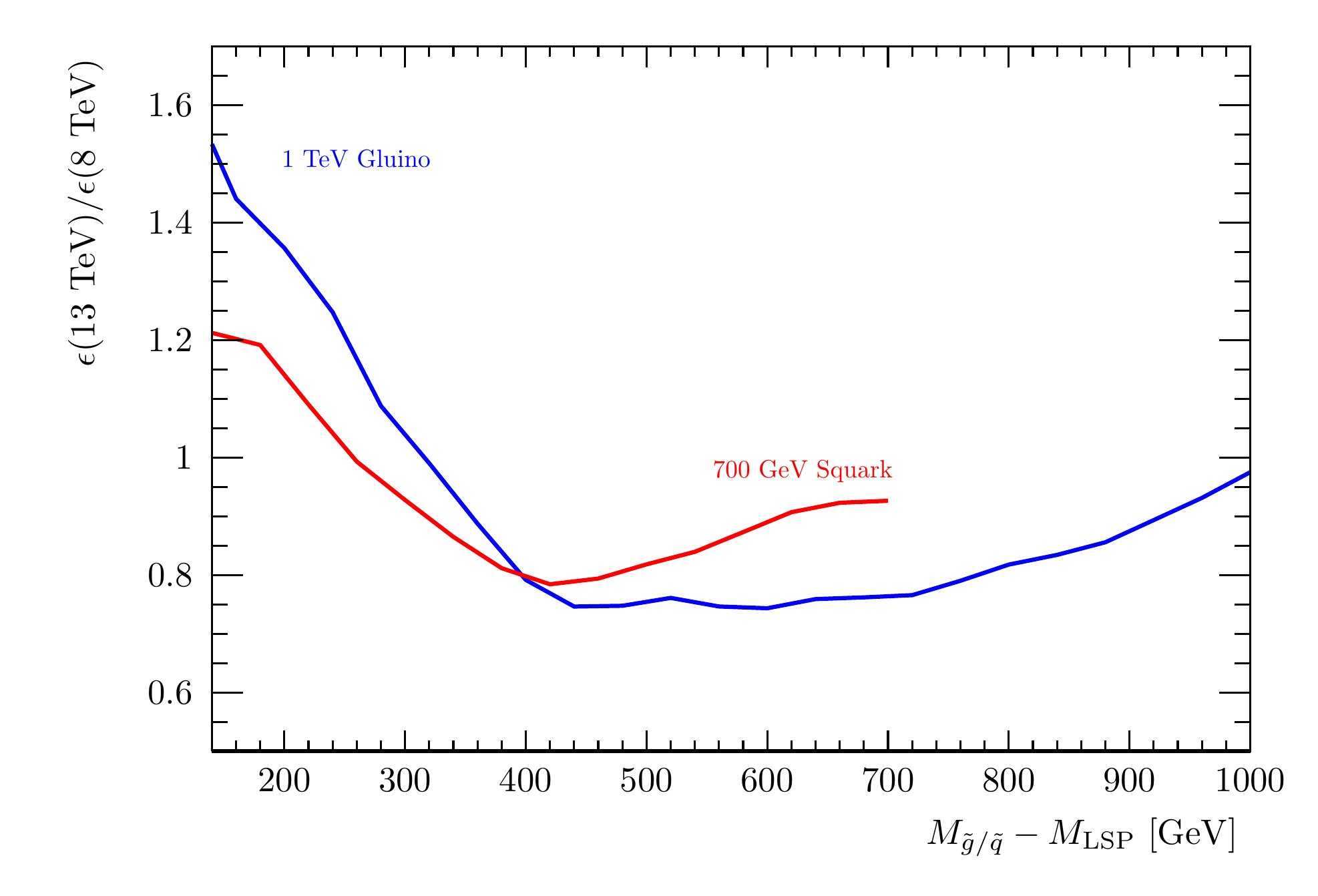}}
\caption{ \label{fig:cs_ratio}
(a):
The ratio of the cross sections at the 13/8 TeV LHC.
On the solid lines, the cross sections at $\sqrt{s}=8$ TeV are greater than 1 fb, which will be the least value to explain the ATLAS8 $Z$+MET excess.
In the case of the 1st/2nd generation squark productions, the line gets closer to the stop line as $m_{\tilde g}\to \infty$.
(b): The ratio of the acceptance rates at the 13/8 TeV LHC.
The blue (red) line shows the 1 TeV gluino (700 GeV squark) case.
Here we assume the gluino decay chain $\tilde g \to g (X_2 \to Z X_1)$ and the squark decay chain $\tilde q \to q (X_2 \to Z X_1)$, where $X_1$ ($X_2$) is the (N)LSP and $M_{X_2} - M_{X_1}=100$ GeV.
}
\end{figure}

Assuming that the excess of the ATLAS Run 2 comes from the same new physics responsible for the Run 1 excess, the observed visible cross section at ATLAS Run 2 is
\begin{align}
(\epsilon\sigma)_{13\text{TeV}}^\text{NP,obs} = 3.3 \pm 1.6~\text{fb}. \label{eqn:visiblexsec13TeV}
\end{align}
We define the observed ratio of the visible cross sections at ATLAS
\begin{align}
R^\text{obs} \equiv \frac{(\epsilon\sigma)_{13\text{TeV}}^\text{NP,obs}}{(\epsilon\sigma)_{8\text{TeV}}^\text{NP,obs}} = 3.7^{+2.7}_{-1.8}.
\end{align}
As we will see later, many proposed models of new physics predict a value of $R$ larger than this observed value. As a result, a large region of the parameter space is typically disfavoured. To understand this, we plot in Fig.~\ref{subfig:cross_ratio} the ratio of the cross sections at the 13/8 TeV LHC for typical colored heavy particle productions. We also show in Fig.~\ref{subfig:acc_ratio} that the ratio of the acceptance rates $(\epsilon)_{13\text{TeV}}/(\epsilon)_{8\text{TeV}}$ is typically of order 1. In Fig.~\ref{subfig:acc_ratio} we assume the gluino decay chain $\tilde g \to g (X_2 \to Z X_1)$ and the squark decay chain $\tilde q \to q (X_2 \to Z X_1)$, where $X_1$ and $X_2$ are the lightest SUSY particle (LSP) and the next-to-LSP (NLSP) respectively, with a mass gap $M_{X_2} - M_{X_1}=100$ GeV. Combining Figs.~\ref{subfig:cross_ratio} and~\ref{subfig:acc_ratio}, we see that the predicted $R$ is generically larger than $R^\text{obs}$. However, we also see that a lower parent particle mass will lead to a smaller $R$, and $R$ is also relatively small if the parent particles are produced by valence quarks (\textit{e.g.} the heavy gluon model or $T$-channel squark production to be discussed later in this paper). Therefore, light parent particles and/or production via valence quarks are favoured for the best fit of the 13/8 TeV ATLAS excesses.

The rest of this paper is organized as follows. In section~\ref{sec:models}, we review various new physics models proposed to explain the 8 TeV ATLAS $Z$+MET excess. We then reduce these models into simplified models in section~\ref{sec:13TeVLHC}, and check the consistency between the 13/8 TeV LHC results. Section~\ref{sec:discussion} is devoted to discussions.

\section{Models for ATLAS Excess}\label{sec:models}

To explain the 8 TeV ATLAS $Z$+MET excess, a model of new physics must satisfy the following three conditions:
\begin{enumerate}
 \item Having a substantial production cross section of the parent particle.
 \item The parent particle having a large decay branching fraction into the $Z$ bosons.
 \item Satisfying the constraints from searches other than the $Z$+MET, such as the jets+MET+zero-lepton search.
\end{enumerate}

To satisfy Condition 1, the parent particle is usually assumed to be produced through the  strong interaction. This is because the electroweak (EW) production requires light parent particles in order to have an adequate cross section, but such a light spectrum cannot survive the event selection cuts, in particular the $H_{\text{T}}$ cut~\cite{Barenboim:2015afa}.

Conditions 1 and 2 together guarantee a substantial production rate of the $Z$ boson, which is necessary in explaining the $Z$+MET excess. For this purpose alone, it seems that one could remove Condition 2, and compensate it by further improving Condition 1, that is by having a larger production cross section of the parent particle. But this generically would not work, due to Condition 3. Therefore, Condition 2 is quite crucial in explaining the $Z$+MET excess. However, in many cases, the $\text{SU}(2)_L$ gauge invariance will make the branching fractions into the $W$ bosons or the Higgs bosons ($h$) comparable to that into the $Z$ bosons. This will violate Condition 2.

To overcome this difficulty, General Gauge Mediation (GGM) models~\cite{Meade:2008wd, Buican:2008ws} with the neutralino NLSP and a very light gravitino LSP can be considered~\cite{Barenboim:2015afa, Allanach:2015xga}. In this case, because the interactions between the SUSY SM particles and the gravitino are tiny, all the decay chains contain the decay of the lightest neutralino into the gravitino. Tuning the parameters of the neutralino sector, one can enhance the branching fraction of the lightest neutralino into the $Z$ boson. Nonetheless, because the gravitino LSP is almost massless in this type of models, there are typically large MET and jet activities, which is in tension with Condition 3~\cite{Allanach:2015xga}. As we will see, the 13 TeV searches provide stronger constraints on the GGM models. In addition, in both 8 TeV and 13 TeV excesses, the jet activity and momentum of the $Z$ boson are small in the signal regions~\cite{Aad:2015wqa, Aad:2015wqa,ATLAS13TeV}. If we take these features into account seriously, the GGM cases may be disfavoured.

Another way of enhancing the branching fraction into the $Z$ boson is to have a compressed mass spectrum. If the mass gaps between new physics particles are of the order of the EW scale, we can tune $\text{SU}(2)_L$ breaking masses so that the branching fraction into the $Z$ boson is selectively enhanced. For instance, consider the decay of the second lightest neutralino into the lightest one $\tilde \chi^0_2 \to \tilde \chi^0_1$. If $ M_Z < M_{\tilde \chi^0_2} - M_{\tilde \chi^0_1} < M_h $, then the branching fraction into the $Z$ boson can dominate over that into $h$. Such a compressed mass spectrum also has an advantage for Condition 3. Therefore many proposed models have a compressed mass spectrum.

In the following, we classify the models proposed to explain the ATLAS $Z$+MET excess into three categories: (1) SUSY models with the gluino production, (2) SUSY models with the squark production, and (3) a non-SUSY model. As mentioned before, the EW production is not considered, since it cannot produce enough events that survive the event selection cuts.

\begin{figure}[h!]
 \centering
 \subcaptionbox{gluino 3-body decay \label{subfig:3-body}}{\includegraphics[width=0.32\textwidth]{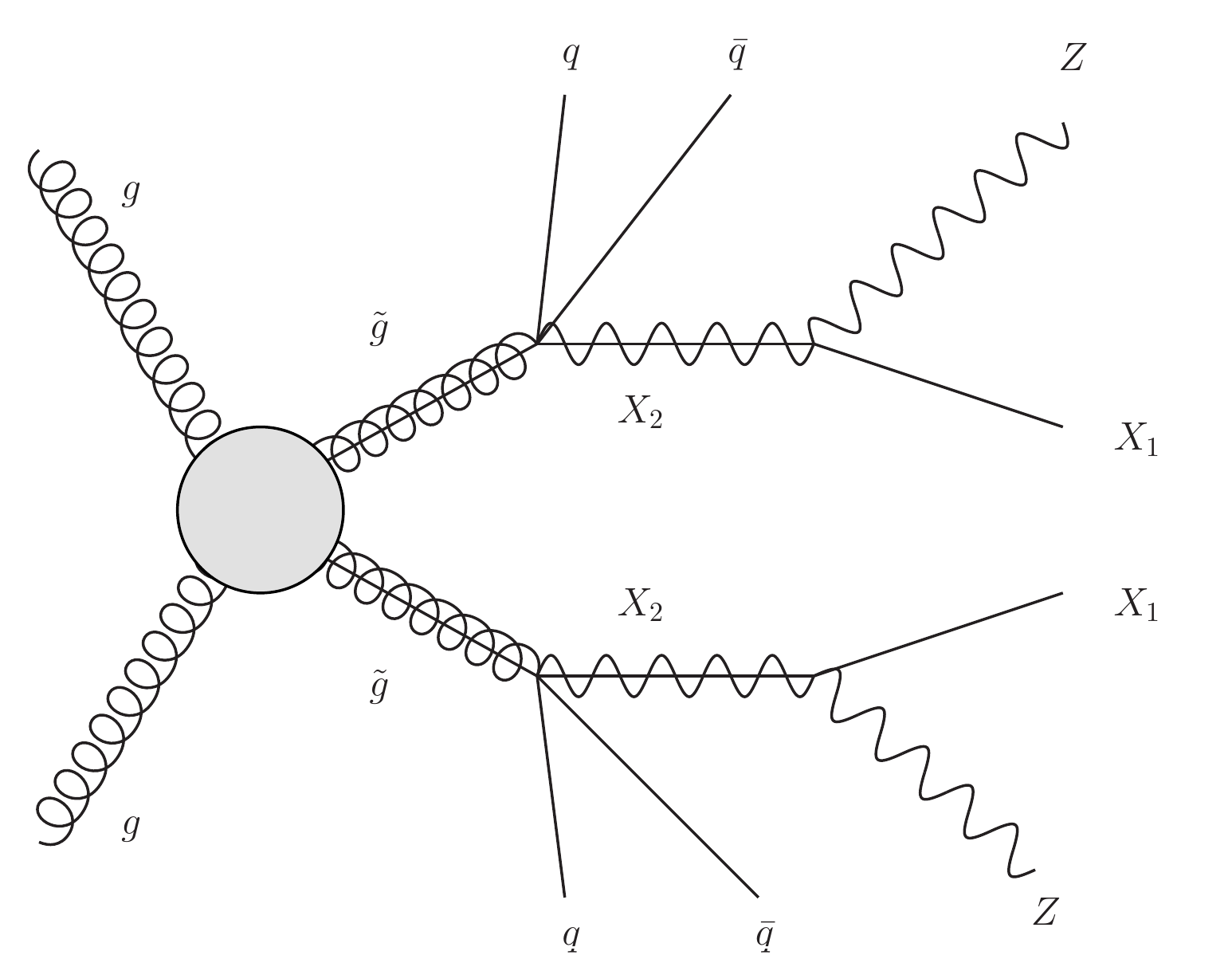}}
    \subcaptionbox{gluino 2-body decay \label{subfig:2-body}}{\includegraphics[width=0.32\textwidth]{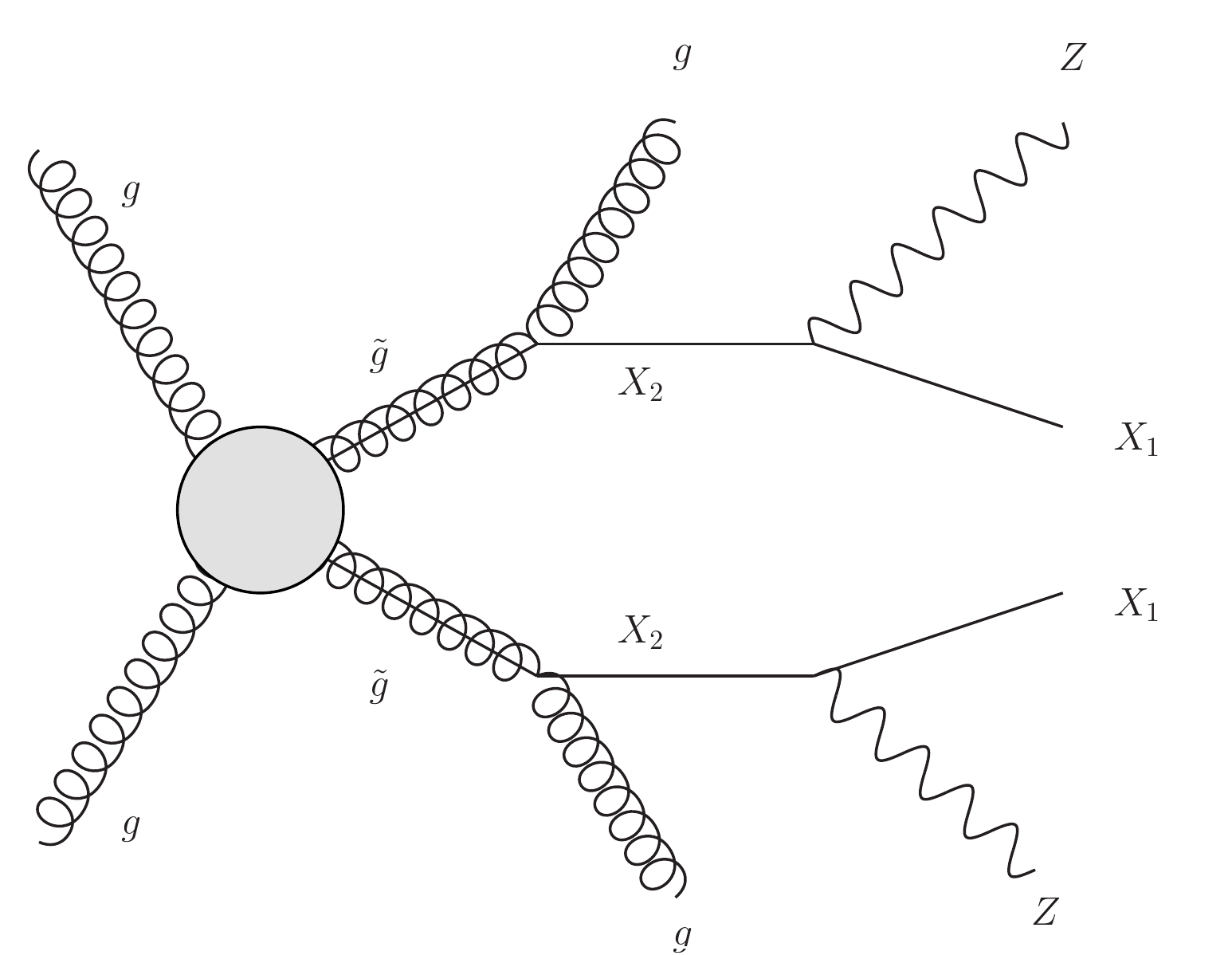}}
      \subcaptionbox{gluino decay via light sbottom \label{subfig:sbottom}}{\includegraphics[width=0.32\textwidth]{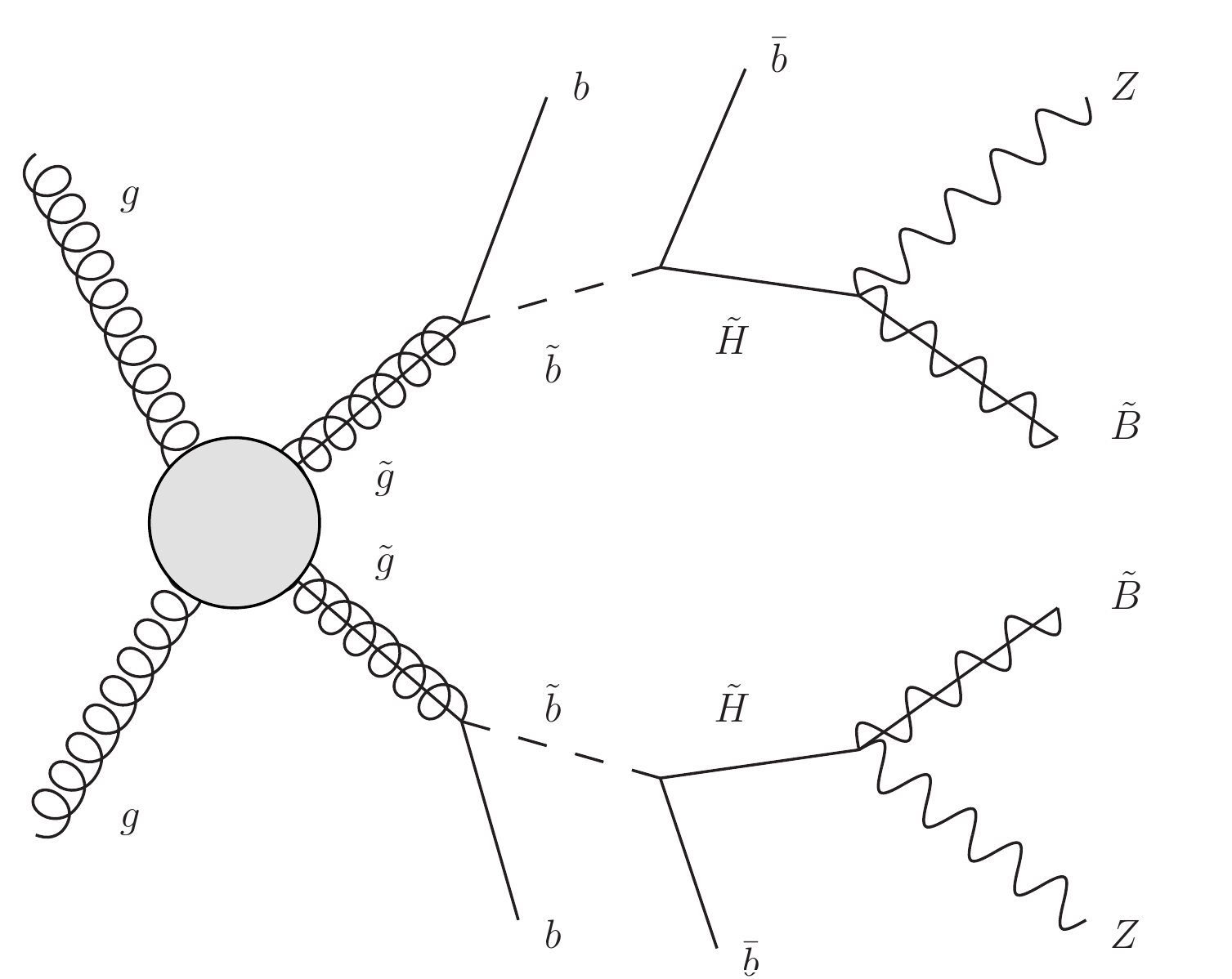}}
          \subcaptionbox{squark pair \label{subfig:squark}}{\includegraphics[width=0.32\textwidth]{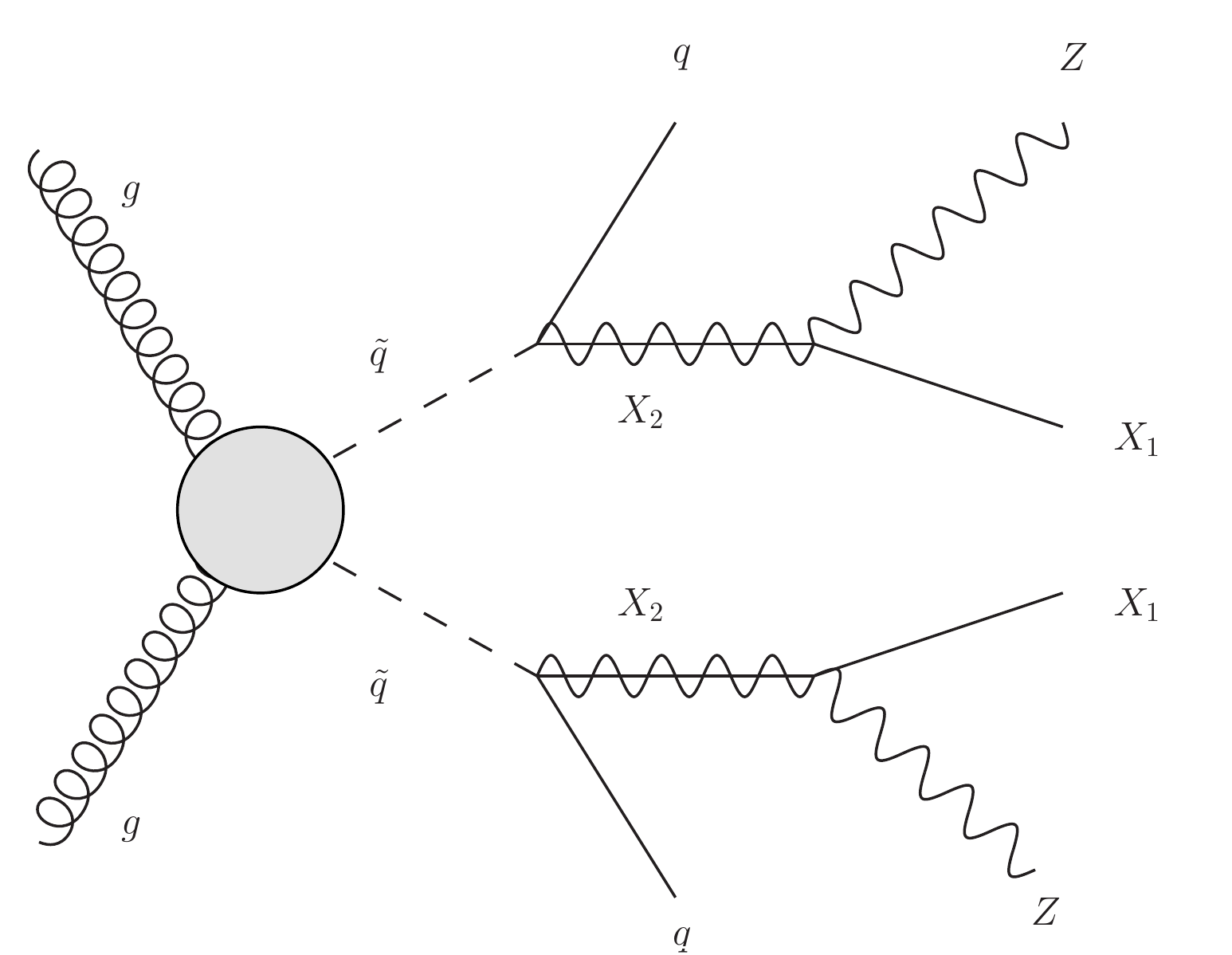}}
    \subcaptionbox{mixed-stop \label{subfig:mixed-stop}}{\includegraphics[width=0.32\textwidth]{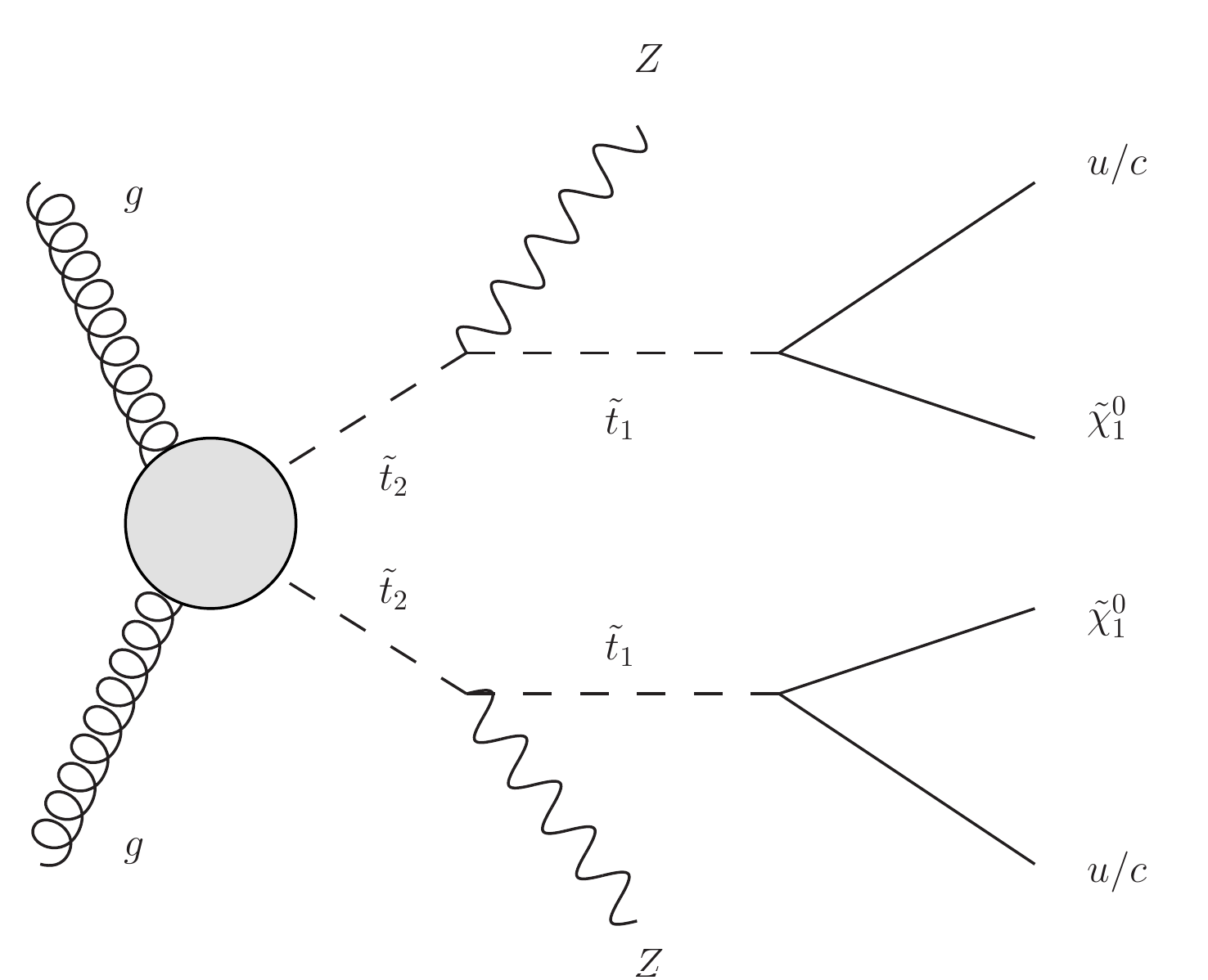}}
      \subcaptionbox{heavy gluon \label{subfig:CHRS}}{\includegraphics[width=0.32\textwidth]{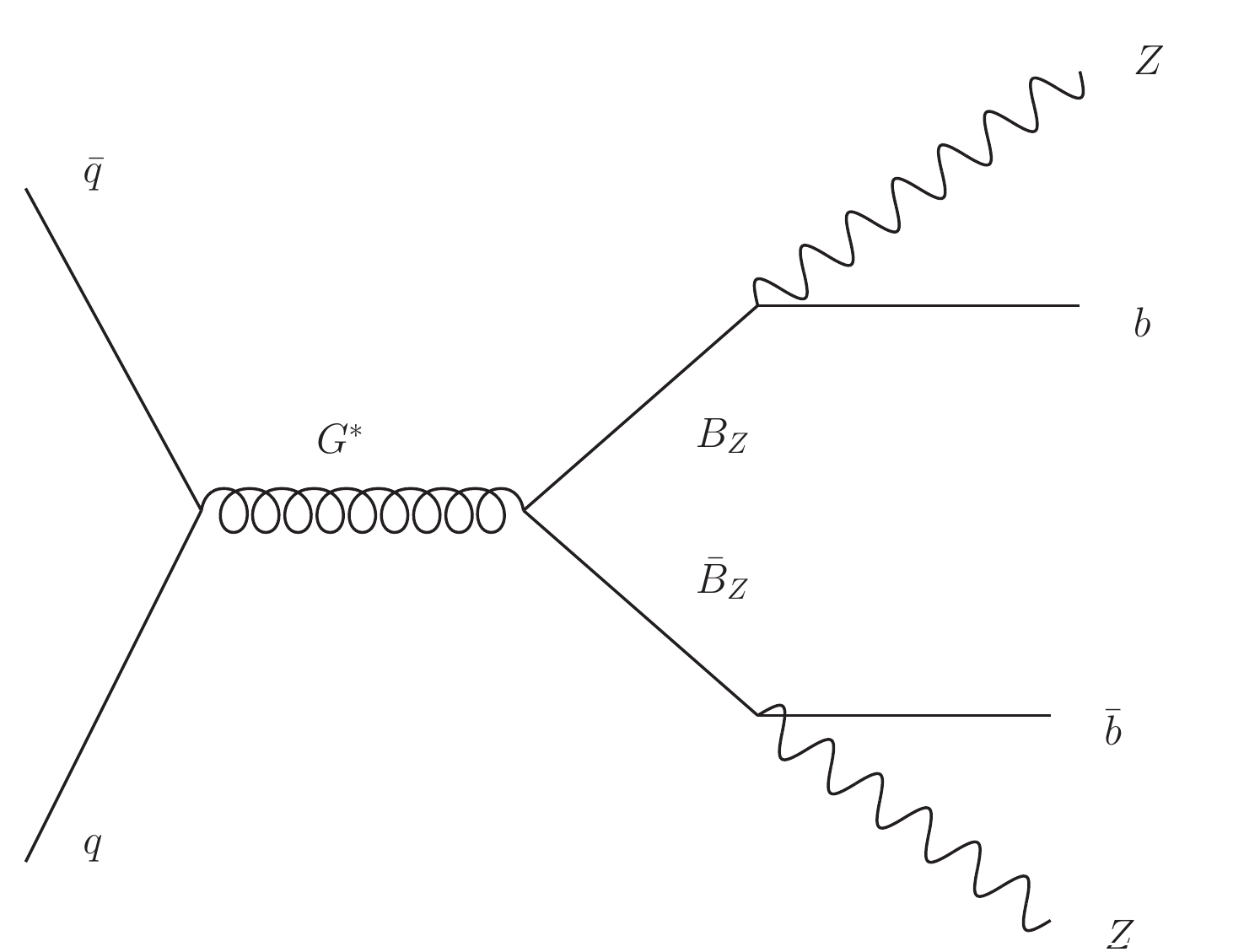}}
  \caption{Typical decay chains with the $Z$ boson emissions in models reviewed in section~\ref{sec:models}.  One of the typical production processes is also depicted in each case.
  }\label{fig:diagrams}
\end{figure}

\subsection{Gluino production}

This category is subdivided according to the decay modes of the gluino.

\subsubsection*{Quark-antiquark emission}

The first gluino decay mode is into a quark-antiquark pair and a neutralino, which subsequently decays into a $Z$ boson and a neutral LSP. This decay chain $\tilde{g} \to q \bar{q} X_2 \to q \bar{q} Z X_1$, with $X_1$ ($X_2$) denoting the neutral (N)LSP, is shown in Fig.~\ref{subfig:3-body}. In the literature, (1) the GGM model~\cite{Barenboim:2015afa, Allanach:2015xga} and (2) the NMSSM models with the gluino production~\cite{Ellwanger:2015hva, Cao:2015ara, Harigaya:2015pma} belong to this case. In the case (1), $X_2 = \tilde{\chi}_1^0$ (the lightest neutralino) and $X_1 = \tilde{G}$ (the gravitino), and one can tune the SUSY SM parameters so that the branching fraction of $\tilde{\chi}_1^0$ to $Z$ and $\tilde{G}$ is close to $100\%$. The direct decay of ${\tilde g}$ into $\tilde{G}$ is suppressed as the interaction between the gravitino and SUSY SM particles is tiny. Because the LSP is nearly massless, this case is strongly constrained by the jets+MET+zero-lepton and multi-lepton searches at the LHC Run 1~\cite{Allanach:2015xga}. In the case (2), $X_2 \simeq \tilde{B}$~\cite{Cao:2015ara} or $X_2 \simeq \tilde{H}$~\cite{Harigaya:2015pma} and $X_1\simeq \tilde{S}$ (the singlino). The direct decay of the gluino into the singlino is suppressed by the smallness of the coupling. In this case, the LSP is massive and the constraints from the jets+MET+zero-lepton search can be relaxed.

\subsubsection*{Gluon emission}

The second gluino decay mode is into a neutralino and a gluon via a top-stop loop. The decay chain is $\tilde{g} \to g X_{2} \to g Z X_1$, as shown in Fig.~\ref{subfig:2-body}. Examples of this category include the mini-split SUSY \cite{Wells:2003tf,*Wells:2004di, ArkaniHamed:2004fb, *Giudice:2004tc, *ArkaniHamed:2004yi,Hall:2011jd, *Hall:2012zp, *Nomura:2014asa, Ibe:2011aa, *Ibe:2012hu, Arvanitaki:2012ps, ArkaniHamed:2012gw} scenario~\cite{Lu:2015wwa}. If $X_2 \simeq \tilde{H}_{u}$, $X_1 \simeq \tilde{B}$ with compressed mass spectrum and heavy (${O}(10^{1-3})$ TeV) squarks, the branching fraction of $\tilde{g} \to g X_{2}$ is enhanced and can be a dominant mode~\cite{Toharia:2005gm, Gambino:2005eh, Sato:2012xf, *Sato:2013bta}. The branching fraction of $X_2\to Z X_1$ is almost 100\% if $M_Z<M_{X_2}-M_{X_1}<M_h$. Another example is a goldstini~\cite{Cheung:2010mc} scenario~\cite{Liew:2015hsa}. In this case, $X_2 \simeq \tilde{B}$ and $X_1 \simeq\tilde{G}'$, where $\tilde{G}'$ is a (massive) pseudo-goldstino. The direct decay of the gluino into the pseudo-goldstino is prevented by the smallness of the interaction. The branching fraction of $\tilde{g} \to g X_{2}$ is enhanced when the higgsinos and winos are significantly heavier than the bino. The branching fraction of $X_2\to Z X_1$ can get to almost $100\%$ if $M_Z<M_{X_2}-M_{X_1}<M_h$.

\subsubsection*{Decay into third generation squark}

The third mode of the gluino decay is that into a lighter squark plus a quark. Due to the effect of the renormalization group equations, the third generation squarks (stops and sbottoms) are likely to be lighter than the other squarks. The light sbottoms/stops scenario is considered in Ref.~\cite{Kobakhidze:2015dra}. The stop/sbottom strongly couples  to the higgsinos, so higgsino is taken as the NLSP and the bino is considered as the LSP. If the mass difference between the gluino and the stop/sbottom is small, the gluino mainly decays into the sbottom for the kinematic reason. This decay chain $\tilde{g} \to b \tilde{b} \to b \bar{b} X_2 \to b \bar{b} Z X_1$, with $X_2 \simeq \tilde{H}_d$ and $X_1 \simeq \tilde{B}$, is shown in Fig.~\ref{subfig:sbottom}. This scenario is severely constrained by the $b$-jets search. For example, only a corner of parameter space (degenerate region) can marginally explain the ATLAS $Z$+MET excess at 2$\sigma$ level~\cite{Kobakhidze:2015dra} while avoiding the ATLAS $\ge3$ $b$-jets + MET constraint~\cite{Aad:2014lra}.

\subsection{Squark production}

The next category is the squark-pair production. This category is subdivided according to the flavor of the produced squarks.

\subsubsection*{Squarks of the first and second generations}

The first case is the production of the 1st/2nd generation squarks. Since the 1st/2nd generation squarks produce less top and bottom quarks in their decay chains, compared to the third generation, they are favoured by the various SUSY searches. The decay chain is $\tilde{q} \to q X_2 \to q Z X_1$, as shown in Fig.~\ref{subfig:squark}. Examples include the NMSSM model with squark production~\cite{Cao:2015zya} and the type of spectrum found by the dedicated parameter scan in the phenomenological MSSM with 19 parameters~\cite{Cahill-Rowley:2015cha}. In the former, one has $X_2 \simeq \tilde{B}$ and $X_1 \simeq \tilde{S}$, with $\tilde{S}$ being the singlino. In the latter, $X_2 \simeq \tilde{B}$ and $X_1 \simeq \tilde{H}$.  In this case, the branching fraction of the bino emitting a $Z$ is not close to 1, and non-negligible amount of $h$ and $W$ are also produced.

\subsubsection*{Mixed stops}

Stop pair production is also a possible scenario. However, since top quarks deposit characteristic signatures like ($b$-)jets, leptons, and/or missing energy, the stop decay mode $\tilde{t} \to t \tilde{\chi}$ is strongly disfavoured. Actually, the branching fraction of $\tilde{t}_2 \to Z \tilde{t}_1$ can become dominant, if the above decay mode is kinematically forbidden (compressed case), or if left- and right-handed stops mix significantly (split case)~\cite{Collins:2015boa}. As to the ${\tilde t}_1$ decay, assuming the decay mode $\tilde{t}_1 \to t \tilde{\chi}_1^0$ is kinematically forbidden, the dominant decay mode will be a flavor-violating two-body decay $\tilde{t}_1 \to q \tilde{\chi}_1^0$ ($q=u, c$), or a flavor-conserving four-body decay $\tilde{t}_1 \to f f' b \tilde{\chi}_1^0$, where $f f'$ are decay products of the off-shell $W$ boson. Among the four combinations (compressed/split and flavor-conserving/violating), it was found in Ref.~\cite{Collins:2015boa} that the flavor-violating compressed case best explains the ATLAS excess while surviving other constraints. Explicitly, the decay chain is $\tilde{t}_2 \to Z \tilde{t}_1 \to Z q \tilde{\chi}_1^0$, as shown in Fig.~\ref{subfig:mixed-stop}. Another advantage of this decay chain is that it may explain the apparent discrepancy between the ATLAS and CMS results, since the $Z$ boson is produced in the first step of the decay chain, and Jet $Z$ Balance (JZB) is distributed relatively symmetrically around 0~\cite{Collins:2015boa}, which causes contamination in the CMS background. When we scan the parameter space in the following section, we extend the above flavor-violating compressed case to the flavor-violating split case, which has the same decay chain.

\subsection{Non-SUSY model with a heavy gluon and vector-like quarks}

The author of Ref.~\cite{Vignaroli:2015ama} considers an effective theory whose particle contents are those of the SM plus a heavy gluon $G^{*}$ and vector-like quarks. It can be realized in a composite Higgs model~\cite{Kaplan:1983fs} or in a Randall-Sundrum model~\cite{Randall:1999ee}. In this effective theory, a pair of heavy quarks $B_Z$ are produced via the $G^*$ resonance, where $B_Z$ has the same quantum numbers as the SM $b$-quark and decays exclusively into a $b$ quark plus a $Z$ boson. The decay chain is shown in Fig.~\ref{subfig:CHRS}. This model has many parameters, which are constrained by the  vector-like quarks signals~\cite{Vignaroli:2015ama}.

\section{13 TeV LHC Test}\label{sec:13TeVLHC}

\subsection{Simulation Setup}

To study the LHC signals and constraints on the various models discussed in section~\ref{sec:models}, we reduce them into simplified models by the following two simplifications: (1) We take all the branching fractions in the decay chains shown in Fig.~\ref{fig:diagrams} as 100\%, except that of the heavy gluon into a $B_{Z}$ pair, which we calculate following Ref.~\cite{Vignaroli:2015ama}. (2) We also decouple all the irrelevant particles (particles not present in Fig.~\ref{fig:diagrams}) from the processes, except in the 1st/2nd generation squark production scenario (Fig.~\ref{subfig:squark}), for which we consider both cases of decoupled gluino and 2.5 TeV gluino.

In our Monte Carlo simulation of the SUSY models, we generate the simplified models with up to one extra parton in the matrix element using MadGraph 5 v2.1.2~\cite{Alwall:2014hca,*Alwall:2011uj} interfaced to Pythia 6.4.28~\cite{Sjostrand:2006za} and Delphes 3~\cite{deFavereau:2013fsa} (with FastJet incorporated~\cite{Cacciari:2011ma,*Cacciari:2005hq}). We apply the MLM matching~\cite{Alwall:2007fs} with a scale parameter set to a quarter of the parent particle (gluino/squark) mass. The parton distribution functions (PDFs) from CTEQ6L1~\cite{Pumplin:2002vw} are used. The production cross sections are calculated at next-to-leading order (NLO) in the strong coupling constant, adding the resummation of soft gluon
emission at next-to-leading-logarithmic accuracy (NLO+NLL) by using NLL-fast v2.1 (for 8 TeV) and v3.0 (for 13 TeV)~\cite{Beenakker:1996ch,*Kulesza:2008jb,*Kulesza:2009kq,*Beenakker:2009ha,*Beenakker:2011fu,*Beenakker:1997ut,*Beenakker:2010nq,*Beenakker:2011fu}.
For the non-SUSY model considered, we use FeynRules 2.3.13~\cite{Alloul:2013bka} to generate the UFO files \cite{Degrande:2011ua}, which is interfaced with MadGraph. The rest of the simulation setup is the same as for the SUSY models.
We have checked that our simulation setup can well reproduce the new physics results by the ATLAS and CMS collaborations.

\subsection{Relevant Experimental Searches}

Our primary interest in this paper is the $Z$+MET excess. We investigate five different $Z$+MET searches: ``ATLAS8 $Z$+MET''~\cite{Aad:2015wqa}, ``ATLAS13 $Z$+MET''~\cite{ATLAS13TeV}, ``CMS8 $Z$+MET''~\cite{Khachatryan:2015lwa}, ``CMS13 (ATLAS-like)'' and ``CMS13 (SRA-B)''~\cite{CMS:2015bsf}. It is useful to summarize and compare their event selection cuts.

First, all of them share the same cut that there is at least one opposite sign same flavor (OSSF) dilepton pair ($e^+e^-$ or $\mu^+\mu^-$) with the on-$Z$ invariant mass $81<m_{ll}<101$ GeV. Moreover, we consider the three searches ATLAS8 $Z$+MET, ATLAS13 $Z$+MET, and CMS13 (ATLAS-like), which are very similar to each other. They all share the following cuts: the large MET $E_{\text{T}}^{\text{miss}} > 225$~GeV, the large hadronic activity $H_{\text{T}}>600$~GeV, at least two signal jets $n_\text{jets}\ge2$, and a minimum azimuthal separation between them and the missing momentum $\Delta\phi (\text{jet}_{12}, p_{\text{T}}^{\text{miss}})>0.4$ to suppress large fake MET events due to jet mismeasurement. The small differences among these three searches are primarily in the transverse momentum thresholds of the leptons and the signal jets. Specifically, ATLAS8 $Z$+MET requires $p_{\text{T}}^{l_1}>25$~GeV, $p_{\text{T}}^{l_2}>10$-$14$~GeV depending on the trigger used, and $p_{\text{T}}^{\text{jet}}>35$ GeV. ATLAS13 $Z$+MET requires $p_{\text{T}}^{l_1}>50$~GeV, $p_{\text{T}}^{l_2}>25$~GeV, and $p_{\text{T}}^{\text{jet}}>30$ GeV. CMS13 (ATLAS-like) requires $p_{\text{T}}^{l_1, l_2}>20$~GeV, $p_{\text{T}}^{\text{jet}}>35$ GeV, and also a minimum separation $\Delta R\equiv\sqrt{(\Delta\phi)^2 +(\Delta\eta)^2}>0.1$ between the dilepton pair, where $\eta$ is the pseudorapidity.

The other two searches, CMS8 $Z$+MET and CMS13 (SRA-B), are more different from the previous three. The CMS8 $Z$+MET search contains two inclusive signal regions: $n_\text{jets}\ge2$ and $n_\text{jets}\ge3$. Each signal region is further divided into three exclusive bins of $E_\text{T}^\text{miss}$: $100-200$, $200-300$, or $>300$ GeV. The transverse momentum thresholds are $p_{\text{T}}^{l_1, l_2}>20$ GeV and $p_{\text{T}}^\text{jet}>40$ GeV. A separation $\Delta R>0.3$ between the signal leptons is also required. The CMS13 (SRA-B) search contains two exclusive signal regions: $n_\text{jets}=2,3$ (SRA) and $n_\text{jets}\ge4$ (SRB). SRA has a $H_\text{T}$ cut $H_\text{T}>400+p_\text{T}^{l_1}+p_\text{T}^{l_2}$ GeV, while SRB does not. Each of them is further divided into $2\times4=8$ exclusive bins according to $n_{b\text{-jets}}^{}$: $0$ or $\ge1$, and $E_{\text{T}}^{\text{miss}}$: $100-150$, $150-225$, $225-300$, or $>300$ GeV. The transverse momentum thresholds and the dilepton separation requirements are the same as CMS13 (ATLAS-like). Table~\ref{tab:cuts_comparison} summarizes these comparisons in detail.

The models discussed in section~\ref{sec:models} typically have accompanying signals. In particular, the jets+MET+zero-lepton channel can often be quite constraining. Therefore, in addition to the five $Z$+MET searches above, we also investigate the constraints from ``ATLAS8 jets+MET'' search~\cite{Aad:2014wea} and ``ATLAS13 jets+MET'' search~\cite{ATLAS0lep13TeV}. The four-lepton searches~\cite{Aad:2014iza,Chatrchyan:2014aea} require production of two $Z$ bosons, both decaying leptonically. Therefore, these signatures strongly depend on the branching fraction of the parent particle into the $Z$ bosons, which is simplified to $100\%$ in our simplified models except the non-SUSY model. Typically, the small leptonic branching fractions of the $Z$ boson, $\text{BF}(Z\to e^+e^-/\mu^+\mu^-)=6.7\%$~\cite{PDG}, makes the signal strength less significant. However, since these searches focus on the small MET and jet activity events, and thus can constrain models with compressed mass spectra, which have the tiny $Z$+MET and multi-jet+MET acceptance rates. In our simplified models, the four-lepton search signal strengths essentially depend on just the cross section of the parent particle $\sigma_\text{parent}$, and its branching fraction into decay chains involving $Z$ bosons. Roughly speaking, $\sigma_\text{parent} \times \text{BF}_{\text{parent}\to Z}^2  \gtrsim 150$ fb, is excluded. One-lepton searches~\cite{Aad:2015mia, ATLAS1lep13TeV} are less important for the simplified models we consider, due to the second lepton veto cut.

\begin{table}[t]
  \begin{center}
  \caption{Event selection cuts of the $Z$+MET searches.}
  \scriptsize
  \begin{tabular}{|c|c|c|c|c|c|c|c|c|c|c|}
  \hline
  \multicolumn{2}{|c|}{} & $E_{\text{T}}^{\text{miss}}$  & $H_{\text{T}}$ & $n_\text{jets}$ & $n_{b\text{-jets}}^{}$ & $\Delta\phi(\text{jet}_{12},p_\text{T}^\text{miss})$ & $p_\text{T}^{l_1}$ & $p_\text{T}^{l_2}$ & $p_\text{T}^\text{jet}$ & $\Delta R$ \\
  \hline
  \multicolumn{2}{|c|}{ATLAS8 $Z$+MET} & \multirow{3}{*}{$>225$}  & \multirow{3}{*}{$>600$} & \multirow{3}{*}{$\ge2$} & & \multirow{3}{*}{$>0.4$} & $>25$ & $>$10-14 & $>35$ & \\
  \cline{1-2}\cline{8-11}
  \multicolumn{2}{|c|}{ATLAS13 $Z$+MET} & & & & & & $>50$ & $>25$ & $>30$ & \\
  \cline{1-2}\cline{8-11}
  \multicolumn{2}{|c|}{CMS13 (ATLAS-like)} & & & & & & $>20$ & $>20$ & $>35$ & $>0.1$ \\
  \hline
  \multirow{4}{*}{CMS8 $Z$+MET} & \multirow{2}{*}{$\ge2$j} & \multirow{2}{*}{100-200} & & \multirow{2}{*}{$\ge2$} & & & \multirow{4}{*}{$>20$} & \multirow{4}{*}{$>20$} & \multirow{4}{*}{$>40$} & \multirow{4}{*}{$>0.3$} \\
                                & & \multirow{2}{*}{200-300} & & & & & & & & \\
  \cline{2-2}\cline{5-5}
                                & \multirow{2}{*}{$\ge3$j} & \multirow{2}{*}{$>300$} & & \multirow{2}{*}{$\ge3$} & & & & & & \\
                                & & & & & & & & & & \\
  \hline
  \multirow{4}{*}{CMS13 (SRA-B)} & \multirow{2}{*}{SRA} & 100-150 & \multirow{2}{*}{$>400+p_\text{T}^{l_1}+p_\text{T}^{l_2}$} & \multirow{2}{*}{2,3} & & & \multirow{4}{*}{$>20$} & \multirow{4}{*}{$>20$} & \multirow{4}{*}{$>35$} & \multirow{4}{*}{$>0.1$} \\
               & & 150-225 & & & 0 & & & & & \\
  \cline{2-2}\cline{4-5}
               & \multirow{2}{*}{SRB} & 225-300 & & \multirow{2}{*}{$\ge4$} & $\ge1$ & & & & & \\
               & & $>300$ & & & & & & & & \\
  \hline
  \end{tabular}
  \label{tab:cuts_comparison}
  \end{center}
\end{table}

\subsection{Results}

In this section, we list our simulation results of the models discussed in section~\ref{sec:models}. For each simplified model, we show the $1\sigma$ and $2\sigma$ fitting regions of the ATLAS8 $Z$+MET~\cite{Aad:2015wqa} excess, and only the $1\sigma$ fitting region of the ATLAS13 $Z$+MET~\cite{ATLAS13TeV} excess. To determine these fitting regions, we first estimate the visible cross sections, $(\epsilon\sigma)_{8\text{TeV}}^\text{NP}$ for the 8 TeV case and $(\epsilon\sigma)_{13\text{TeV}}^\text{NP}$ for the 13 TeV case. We estimate a $\chi^2$ variable for the total number of the signal events, assuming the SM background uncertainty is Gaussian. We refer to the parameter regions where $\Delta\chi^2<1 \, (4)$  as the $1\sigma$ $(2\sigma)$ fitting regions. Note that we do not show the $2\sigma$ fitting region of the ATLAS13 $Z$+MET excess, because this excess is only $2.2\sigma$ and showing the $2\sigma$ fitting region is pointless. Instead we show the 95\% $\text{CL}_S$ exclusion limit from this search. In addition, the $95\%$ $\text{CL}_S$ exclusion limits from the CMS8 $Z$+MET~\cite{Khachatryan:2015lwa}, CMS13 (ATLAS-like) and CMS13 (SRA-B)~\cite{CMS:2015bsf}, ATLAS8 jets+MET~\cite{Aad:2014wea}, and ATLAS13 jets+MET~\cite{ATLAS0lep13TeV} are also shown.

\subsubsection{SUSY GGM Models}

In this category, we consider the SUSY GGM models, in which the LSP is the gravitino and massless. The parent particles are either a gluino pair or a squark pair, with the decay chain $\tilde g \to q \bar{q} (\text{NLSP} \to Z+\text{LSP})$ (Fig.~\ref{subfig:3-body}) and $\tilde q \to q (\text{NLSP} \to Z+\text{LSP})$ (Fig.~\ref{subfig:squark}),  respectively. In the gluino case, the squarks are assumed to be very heavy and the gluino decays into four light flavor quarks ($u,d,s$, and $c$). In the squark case, $\tilde Q_{L,12}$, $\tilde u_{r,12}$ and $\tilde d_{r,12}$ are light with degenerate masses, and the gluino is assumed to be decoupled. We take the masses of the gluino/squark and the NLSP as free parameters.

\begin{figure}[tbp]
\centering
 \subcaptionbox{GGM with gluino \label{subfig:go3bd_GGM}}{\includegraphics[width=0.47\textwidth]{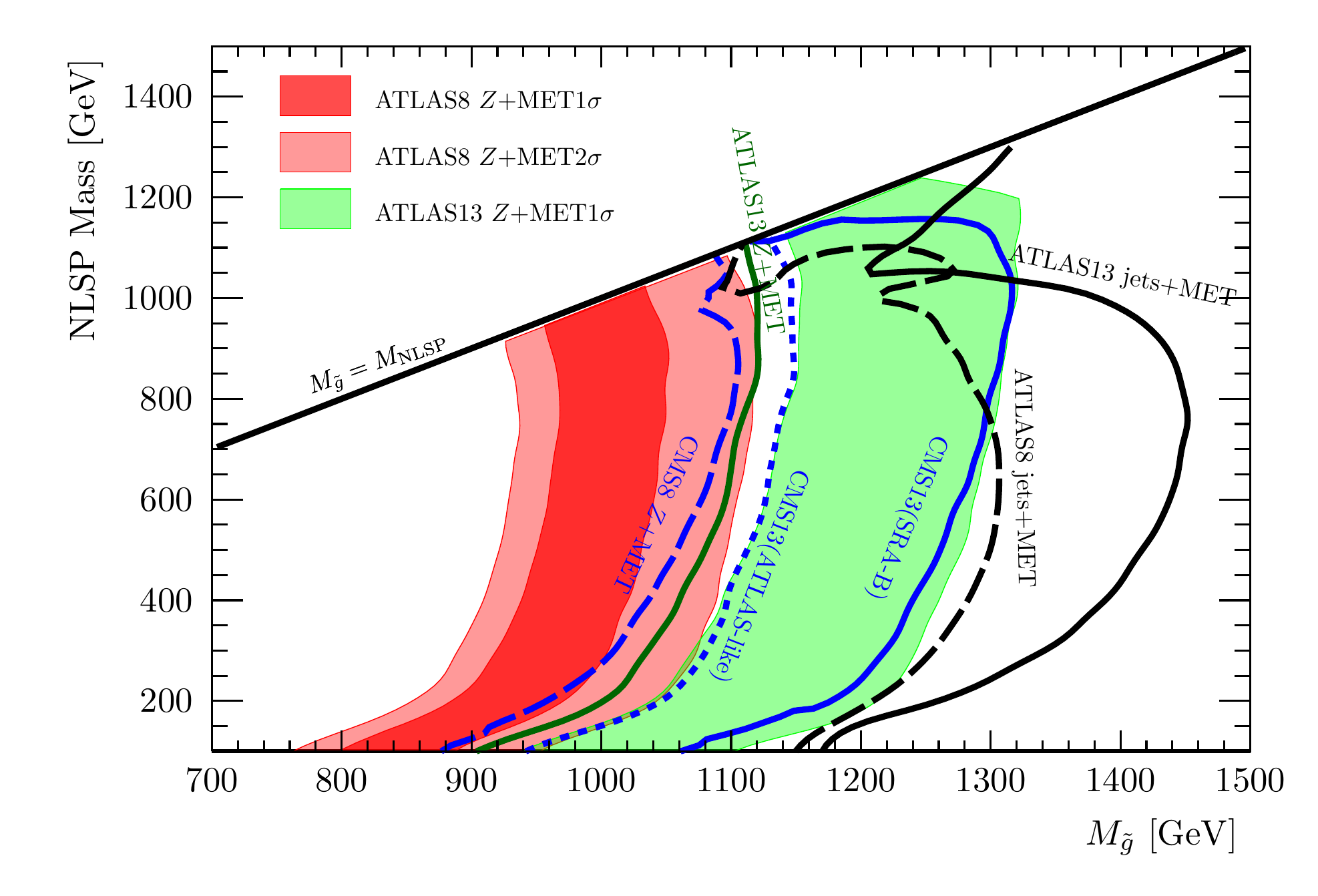}}
 \subcaptionbox{GGM with squark \label{subfig:sb_GGM}}{\includegraphics[width=0.47\textwidth]{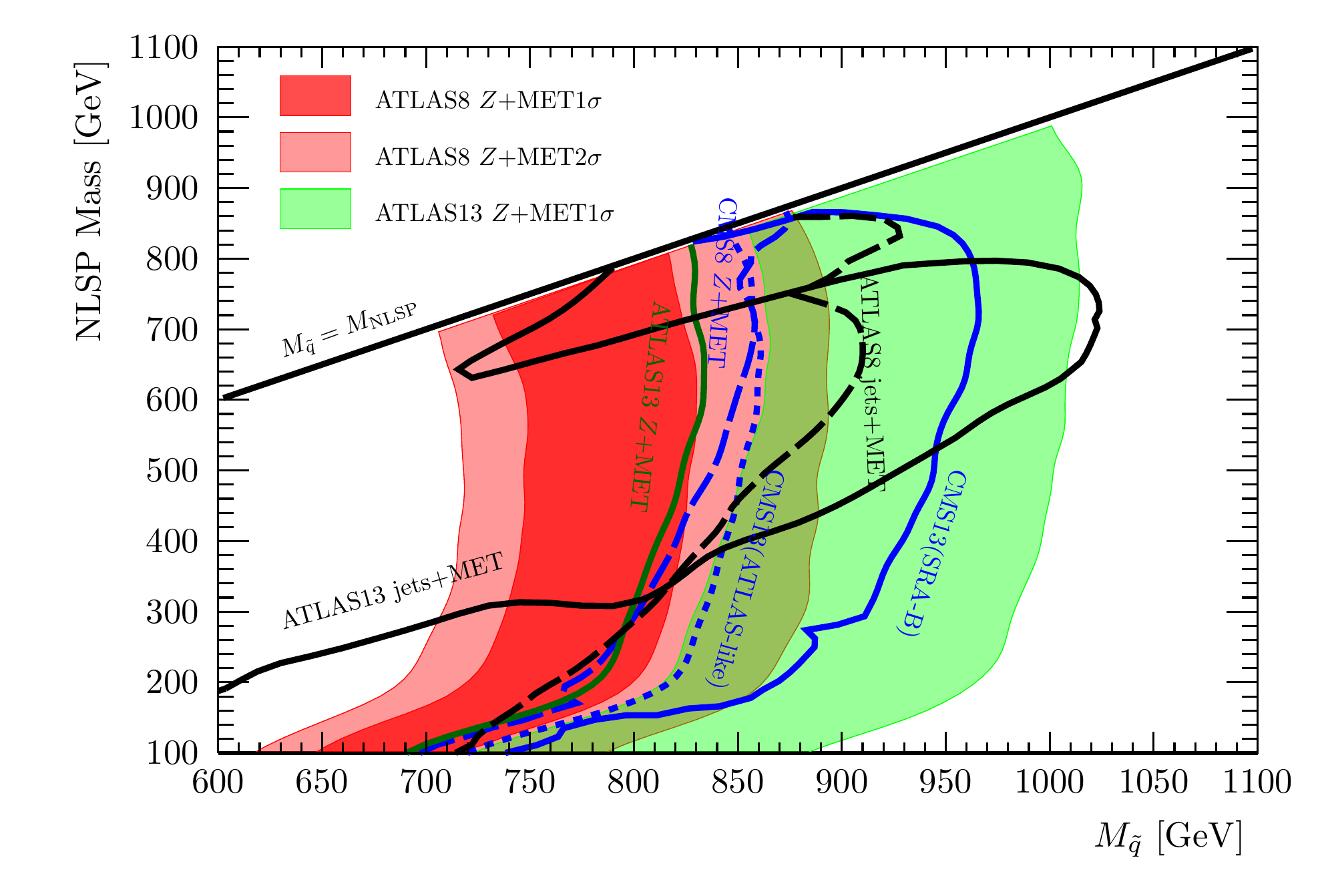}}
\caption{The $1\sigma$ and $2\sigma$ fitting regions of the ATLAS $Z$+MET excess and the  $95\%$ exclusion limits from the various constraints for the simplified GGM models: (a) gluino pair production with the decay chain $\tilde g \to q \bar{q} (\text{NLSP} \to Z+\text{LSP})$ (Fig.~\ref{subfig:3-body}), and (b) squark and anti-squark pair production with the decay chain $\tilde q \to q (\text{NLSP} \to Z+\text{LSP})$ (Fig.~\ref{subfig:squark}). The four-lepton searches exclude the gluino (squark) mass up to about 800 (550) GeV.
}\label{fig:GGM}
\end{figure}

The fitting regions and the exclusion limits are shown in Fig.~\ref{fig:GGM}. We see that there is some tension between ATLAS8 $Z$+MET and ATLAS13 $Z$+MET fitting regions. In addition, the ATLAS jets+MET constraints are quite powerful and exclude a large portion of the fitting regions, especially in the case of the gluino production. The CMS13 (SRA-B) constraints also exclude most of the fitting regions as expected. The four-lepton searches exclude gluino (squark) mass up to about 800 (550) GeV.

\subsubsection{Compressed SUSY Models}

In this category, we consider the compressed SUSY models, in which the masses of the NLSP and the LSP are close. We study several scenarios according to different parent particles and decay chains.

\subsubsection*{Gluino pair production scenario}

The first scenario we consider is the gluino pair production. The gluino decay chain is either  3-body decay (Fig.~\ref{subfig:3-body}), or  2-body decay (Fig.~\ref{subfig:2-body}). In the 3-body decay case, the gluino decays into four light flavor quarks ($u,d,s$, and $c$). We fix the mass gap between the NLSP and the LSP to $100$ GeV, and take the gluino mass and the LSP mass as free parameters.

\begin{figure}[tbp]
\centering
 \subcaptionbox{Gluino 3-body \label{subfig:go3bd}}{\includegraphics[width=0.47\textwidth]{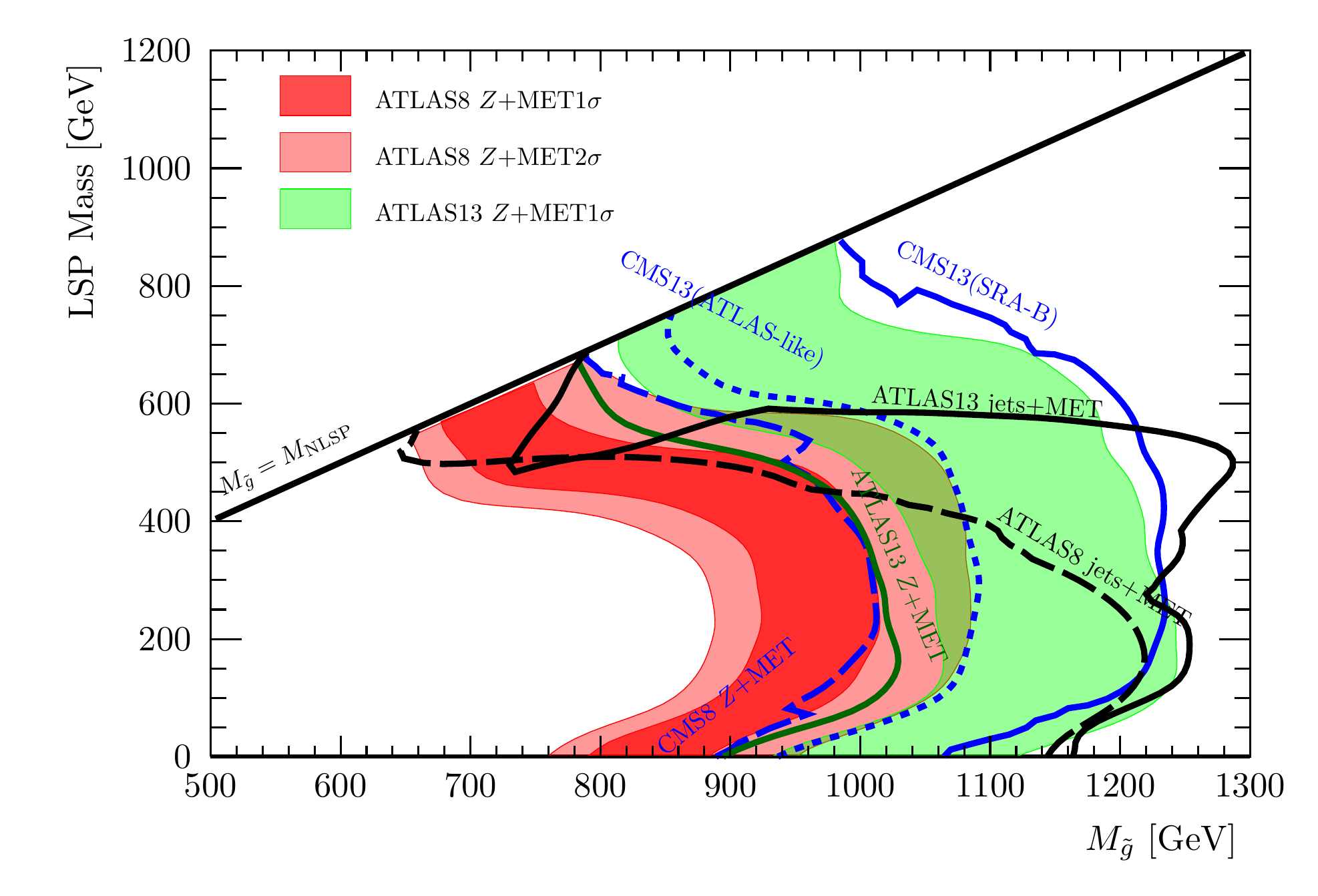}}
 \subcaptionbox{Gluino 2-body \label{subfig:go2bd}}{\includegraphics[width=0.47\textwidth]{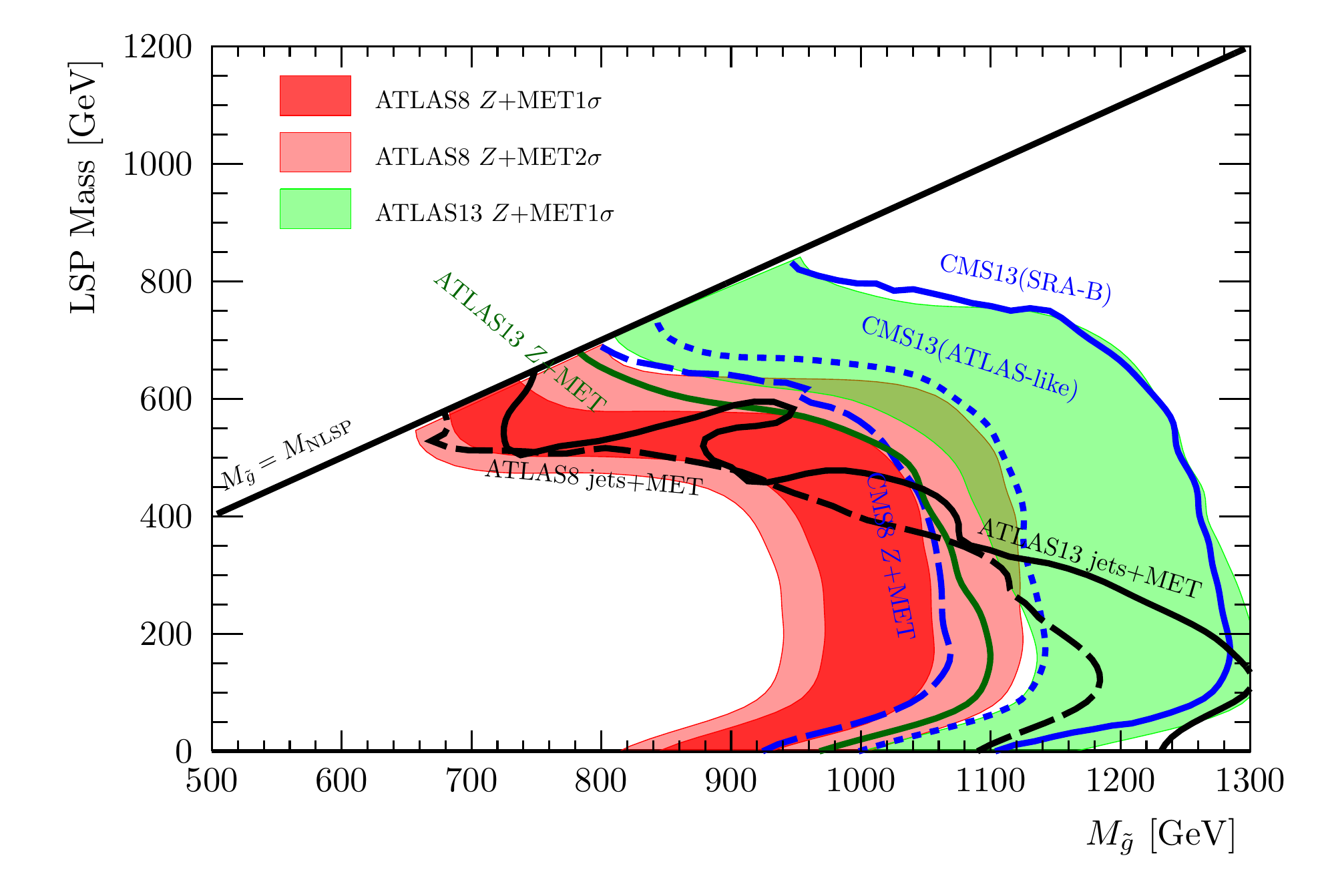}}
\caption{The $1\sigma$ and $2\sigma$ fitting regions of the ATLAS $Z$+MET excess and  the $95\%$ exclusion limits from the various constraints for the simplified compressed SUSY models with gluino pair production: (a) gluino 3-body decay chain $\tilde g \to q \bar{q} (\text{NLSP} \to Z+\text{LSP})$ (Fig.~\ref{subfig:3-body}), and (b) gluino 2-body decay chain $\tilde g \to g (\text{NLSP} \to Z+\text{LSP})$ (Fig.~\ref{subfig:2-body}). The four-lepton searches exclude the gluino mass up to 800 (700) GeV for light (heavy) LSP regions.
}\label{fig:gluino}
\end{figure}

The fitting regions and the exclusion limits are shown in Fig.~\ref{fig:gluino}. We see that the ATLAS jets+MET constraints are less effective, compared to the GGM cases, and the degenerate regime of the fitting region (where the NLSP mass is close to the gluino mass) is consistent with the ATLAS jets+MET searches. Although the tension between ATLAS8 $Z$+MET and ATLAS13 $Z$+MET fitting regions is slightly better than the GGM cases, the ATLAS13 result excludes almost all $1\sigma$ fitting regions for the ATLAS8 $Z$+MET excess. The CMS8/13 constraints have quite severe tension with the ATLAS8 $Z$+MET excess. The four-lepton searches exclude the gluino mass up to 800 (700) GeV for the light (heavy) LSP regions.

\subsubsection*{Squark pair production scenario}

We next consider the squark pair production scenario. In this scenario, the gluino mass is also an important parameter. If the gluino is extremely heavy or if the gluino has the Dirac mass term \cite{Ding:2015jya} and hence is effectively decoupled, then the squark-antisquark pair productions through $S$-channel will dominate. On the other hand, if the gluino is relatively light, then the squark-squark pair productions through $T$-channel will be dominant. Therefore, we consider two cases, one with the decoupled gluino, and the other with the 2.5 TeV gluino. In each case, we fix the mass gap between the NLSP and the LSP to $100$ GeV, and take the squark mass and the LSP mass as free parameters.

\begin{figure}[tbp]
\centering
 \subcaptionbox{squark with decoupled gluino \label{subfig:sb}}{\includegraphics[width=0.47\textwidth]{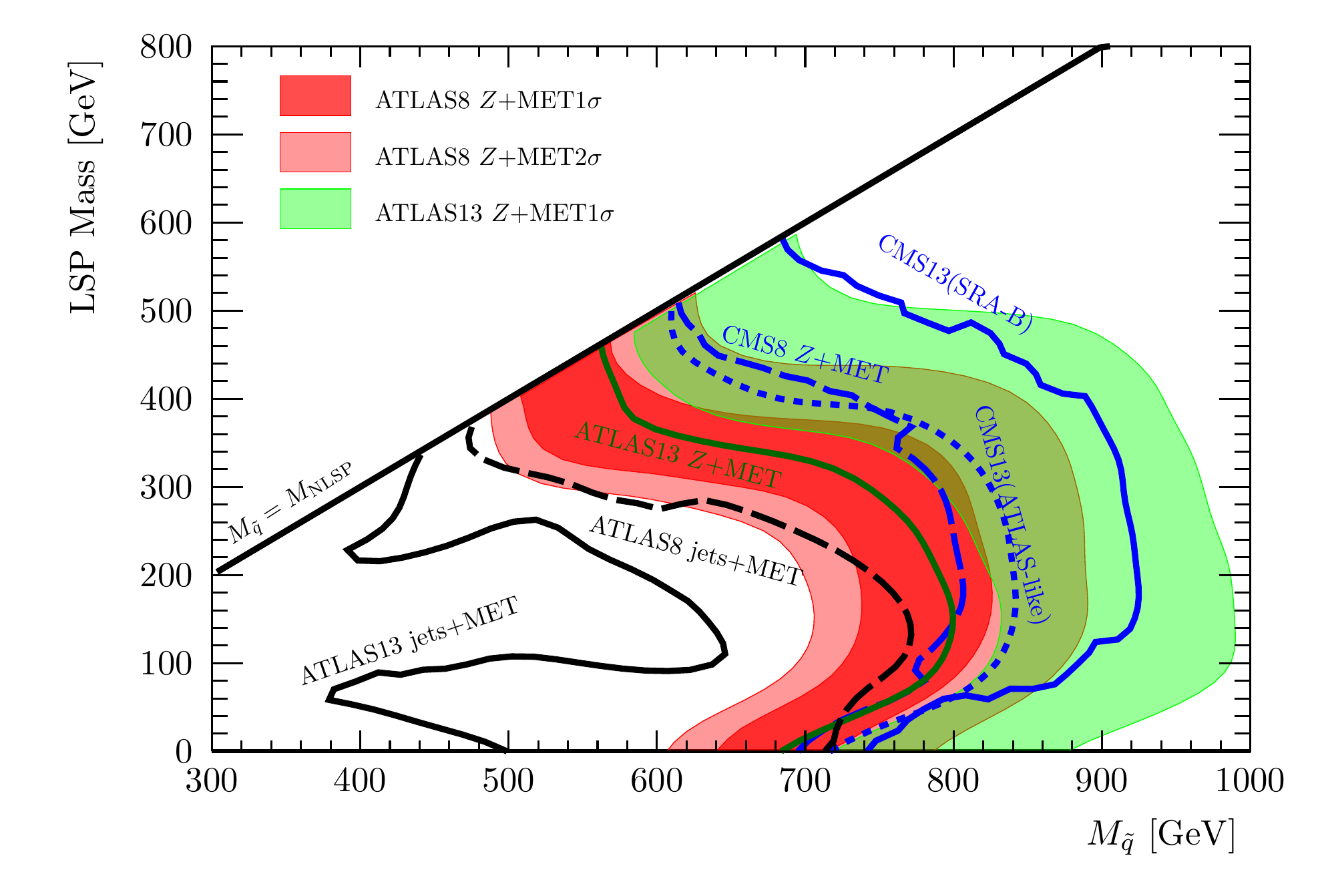}}
 \subcaptionbox{squark with 2.5 TeV gluino \label{subfig:sq_2500}}{\includegraphics[width=0.47\textwidth]{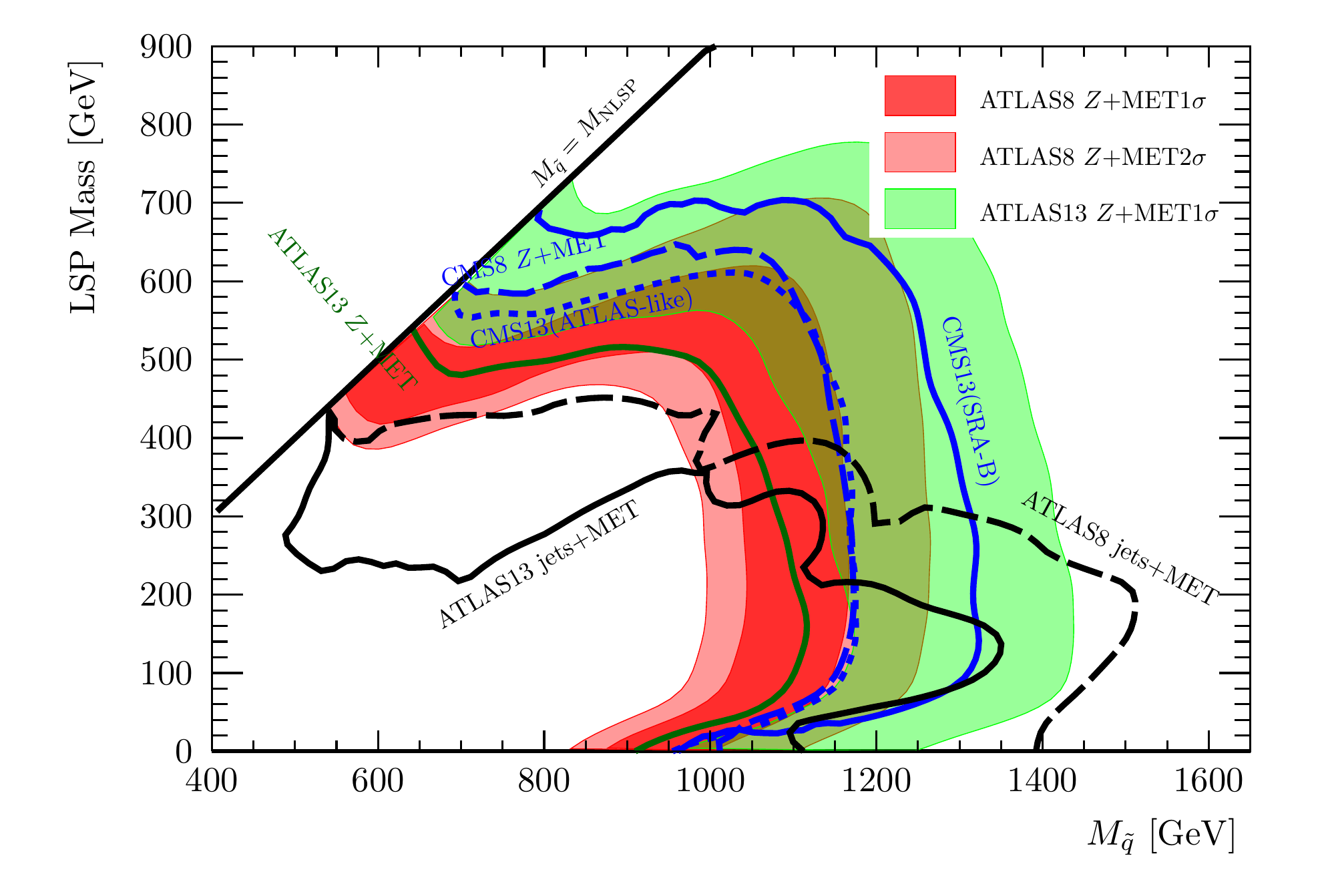}}
\caption{The $1\sigma$ and $2\sigma$ fitting regions of the ATLAS $Z$+MET excess and the  $95\%$ exclusion limits from the various constraints for the simplified compressed SUSY models with the squark pair production (Fig.~\ref{subfig:squark}): (a) squark production with decoupled gluino, and (b) squark production with the $2.5$ TeV gluino.  The four-lepton searches exclude the squark mass up to about 500 (800) GeV for case (a) ((b)).
}\label{fig:squark}
\end{figure}

The fitting regions and the exclusion limits are shown in Fig.~\ref{fig:squark}. We see that similarly to the gluino production scenario, the degenerate regime of the fitting region is consistent with the jets+MET searches. In addition, the tension between the ATLAS8 $Z$+MET and ATLAS13 $Z$+MET fitting regions is also ameliorated. This is because the squarks with the decoupled gluino have the smaller production cross section. On the other hand, the squarks with the light gluino are mainly produced by the valence quarks. In both cases, the $R$ value tends to be small and leads to the better consistency between the 8 TeV and 13 TeV ATLAS results, and a small portion of the fitting region is not excluded by the CMS $Z$+MET and multi-jets+MET constraints. In the decoupled gluino case, the almost massless LSP is relatively favoured, as in the case of the GGM.
In the 2.5 TeV gluino case, the heavy squark with degenerated spectrum is favoured. The four-lepton searches exclude the squark mass up to about 500 (800) GeV in the decoupled (2.5 TeV) gluino case.

\subsubsection*{Scenarios with light third generation squark}

As discussed in section~\ref{sec:models}, some models with the light third generation squark can also explain the ATLAS $Z$+MET excess. These include (1) a gluino pair production with the gluino decay via a light sbottom (Fig.~\ref{subfig:sbottom}), and (2) a stop pair production in the mixed stop model (Fig.~\ref{subfig:mixed-stop}). In scenario (1), we fix the mass gaps $M_\text{NLSP}-M_\text{LSP}=100$ GeV, $M_{\tilde b}-M_\text{NLSP}=150$ GeV, and take the gluino mass and the LSP mass as free parameters. In scenario (2), we fix the mass gap between ${\tilde t}_1$ and the LSP to $25$ GeV, and take the ${\tilde t}_2$ mass and the mass gap $M_{{\tilde t}_2}-M_{{\tilde t}_1}$ as free parameters.

\begin{figure}[tbp]
\centering
 \subcaptionbox{gluino with light sbottom \label{subfig:go_bbbar}}{\includegraphics[width=0.47\textwidth]{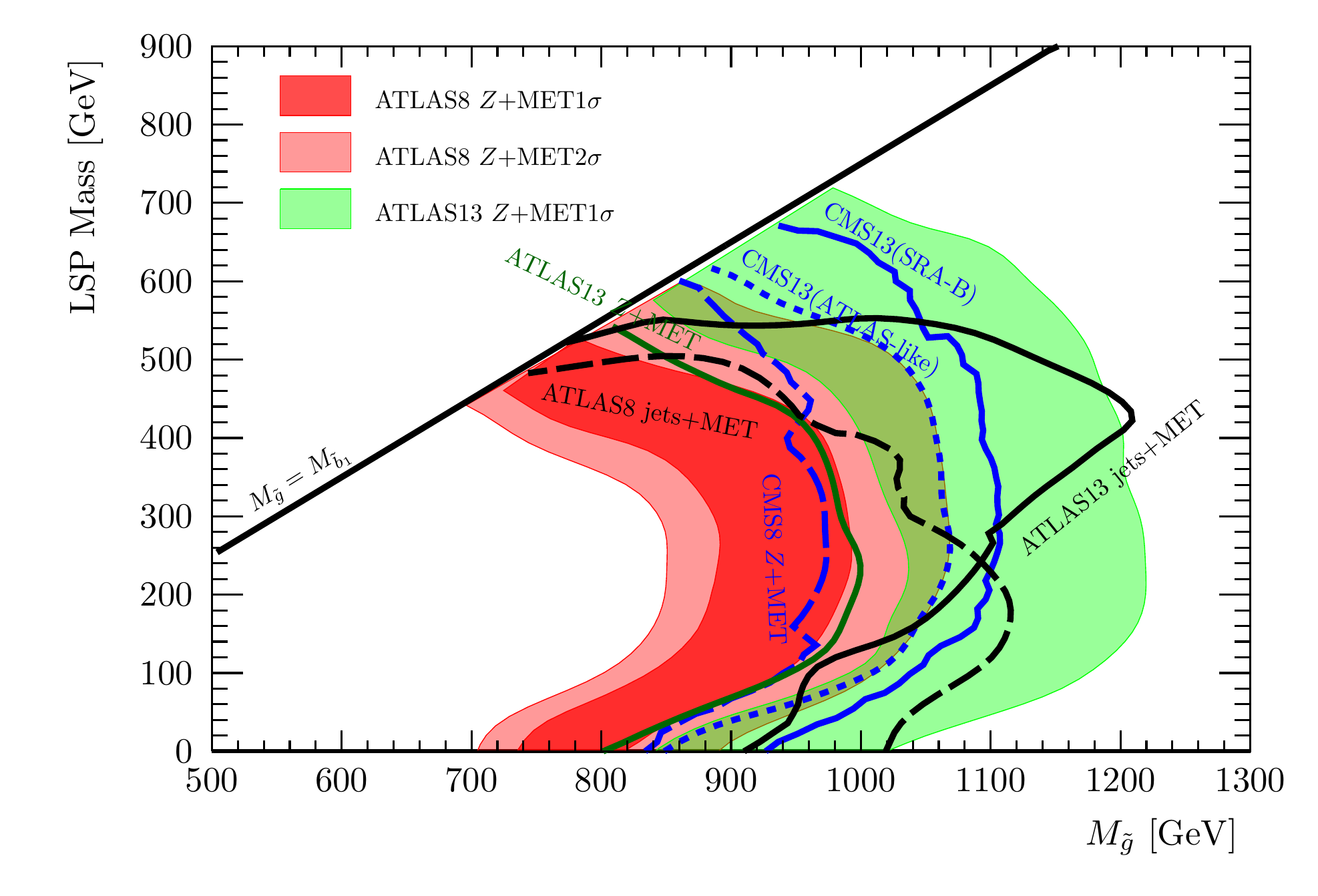}}
 \subcaptionbox{mixed stop \label{subfig:stop}}{\includegraphics[width=0.47\textwidth]{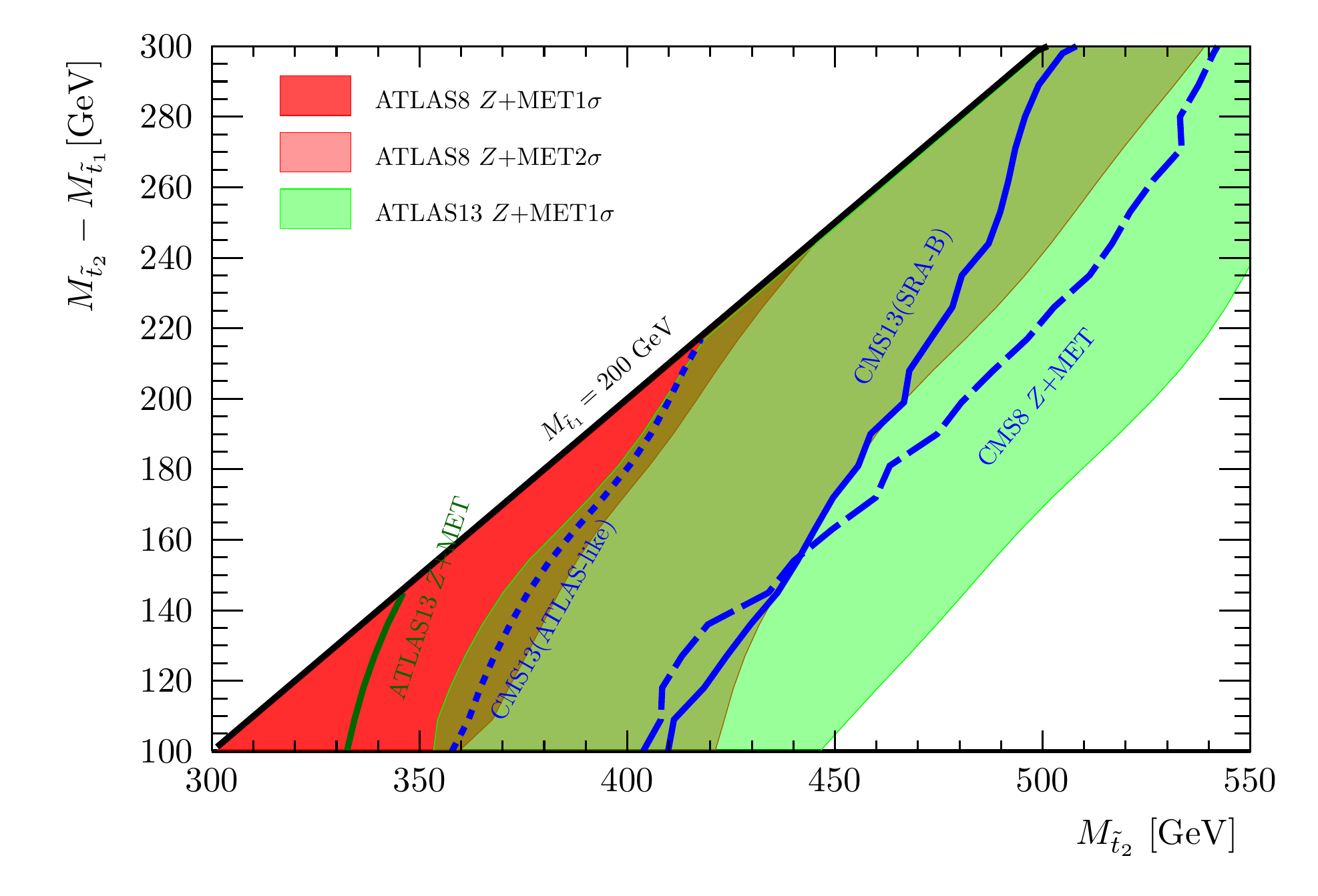}}
\caption{The $1\sigma$ and $2\sigma$ fitting regions of the ATLAS $Z$+MET excess and the  $95\%$ exclusion limits from the various constraints for the simplified compressed SUSY models with light third generation squarks: (a) gluino pair production with decay via the light sbottom (Fig.~\ref{subfig:sbottom}), and (b) the mixed stop case with the stop pair production (Fig.~\ref{subfig:mixed-stop}).
The constraint on the direct production of $\tilde{t}_1$ is $M_{\tilde t_1} \gtrsim 250$ GeV \cite{Aad:2014nra}.
The four-lepton searches exclude the gluino (${\tilde t}_2$) mass up to 750 (400) GeV in the case (a) ((b)).
}\label{fig:third}
\end{figure}

The fitting regions and the exclusion limits are shown in Fig.~\ref{fig:third}. We see that the result of the scenario (1) is qualitatively similar to that of the gluino 3-body decay scenario shown in Fig.~\ref{subfig:go3bd}. This scenario is rich in the $b$-jets signal, and hence subject to the severe $b$-jet search constraints. Moreover, the present ATLAS 8 TeV excess does not prefer too many $b$-jets as seen in Table~\ref{tab:observation}. If we take the $b$-jet number distribution into account, a large portion of the fitting region will be disfavoured. For this simplified model, the four-lepton searches exclude the gluino mass up to 750 GeV.

In the scenario (2), the ATLAS8 $Z$+MET and ATLAS13 $Z$+MET fitting regions can be consistent, and the ATLAS jets+MET constraints are not severe. However, the CMS8 and 13 $Z$+MET constraints are severe. In this analysis, we adopt the background estimations provided by the ATLAS and CMS collaborations. However, as discussed in Ref.~\cite{Collins:2015boa}, the SUSY events in this model can contaminate the background estimations.  In such cases, the tension between the CMS and ATLAS observations may be relaxed. For this simplified model, the four-lepton searches exclude the ${\tilde t}_2$ mass up to 400 GeV.

\subsubsection{Non-SUSY Model}

For the non-SUSY model discussed in section~\ref{sec:models}, we fix the mass of $B_Z$ at 930 GeV, and take the mass of the heavy gluon $G^*$ and the mixing angle $\tan\theta_3$ as free parameters, following Ref.~\cite{Vignaroli:2015ama}.

\begin{figure}[tbp]
\centering
\includegraphics[width=0.47\textwidth]{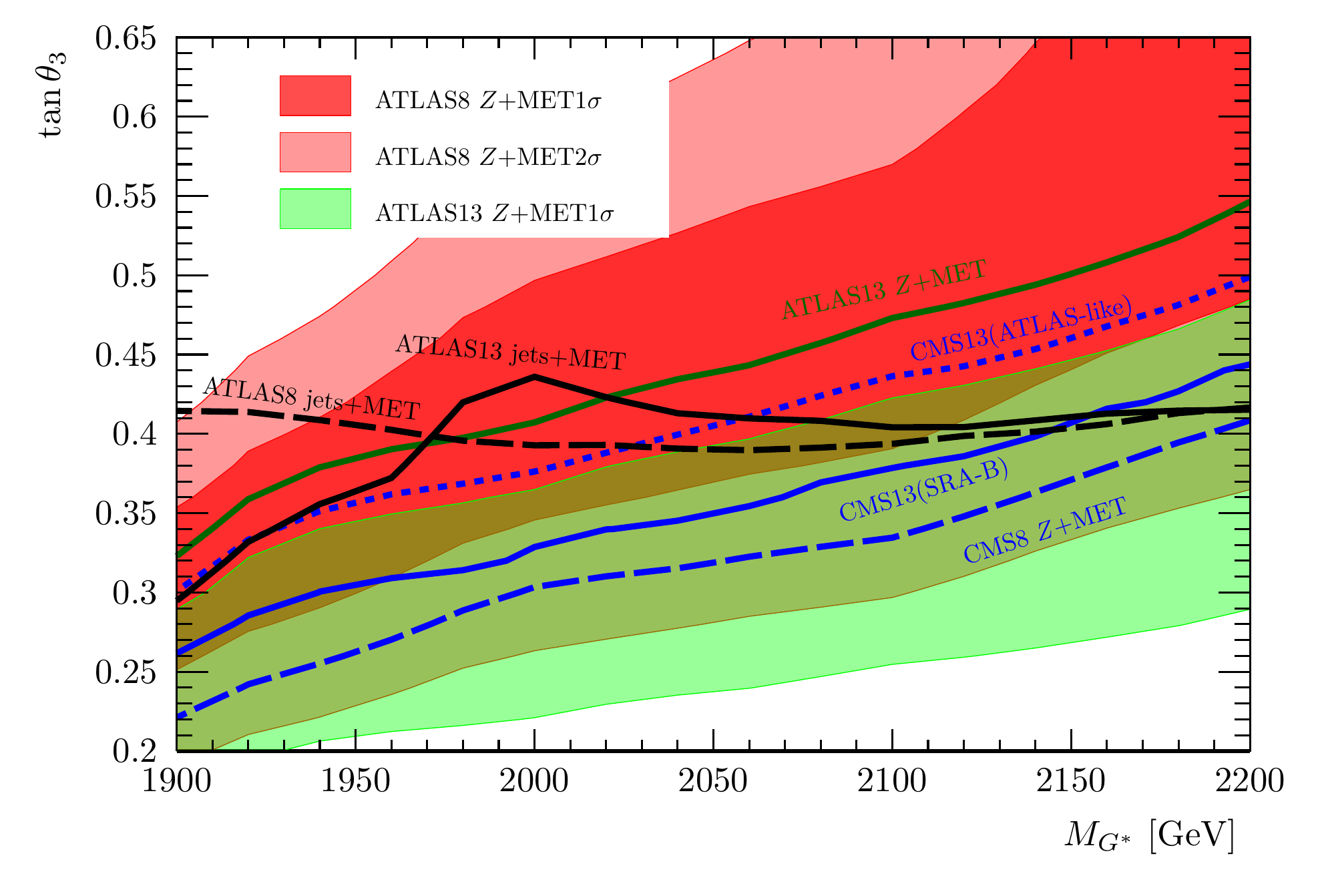}
\caption{$1\sigma$ and $2\sigma$ fitting regions of the ATLAS $Z$+MET excess and $95\%$ exclusion limits from various constraints for the simplified non-SUSY model with a heavy gluon and vector-like quarks (Fig.~\ref{subfig:CHRS}).
}\label{fig:gheavy}
\end{figure}

The fitting regions and the exclusion limits are shown in Fig.~\ref{fig:gheavy}. In the figure, the regions above the exclusion curves are excluded. In this analysis, we do not include the  production modes of the vector-like fermions other than the $B_Z$ fermion nor the SM fermions through the heavy gluon. We see that the tension between the ATLAS jets+MET and the CMS $Z$+MET constraints are relatively mild in the fitting regions of the ATLAS excesses. This scenario is rich in energetic $b$-jet signal, and hence possibly subject to the severe $b$-jet search constraints. Moreover, in this model, the momenta of the $Z$ bosons are larger, since the $Z$ bosons come from the heavy fermion decay $B_Z \to Z b$. Therefore it may conflict with the relatively low momenta of $Z$ bosons in the signal region observed by the ATLAS collaboration.

\subsection{Summary}

Here we summarize the consistency of the ATLAS8 $Z$+MET excess with the other LHC searches we considered.

\subsubsection*{Multi-jet+MET}

If the mass difference between the parent particle and the LSP is large, ATLAS8/13 multi-jet+MET constraints conflict with the ATLAS8 $Z$+MET excess. On the other hand, if the mass spectrum is compressed, these constraints are weak and the current constraints from the ATLAS 8/13 TeV results cannot exclude the ATLAS8 $Z$+MET excess.

\subsubsection*{$Z$+MET by ATLAS13 and CMS13 ATLAS-like channel}

Although the ATLAS13 $Z$+MET search reports a $2.2\sigma$ excess, it can actually constrain some models explaining the ATLAS8 $Z$+MET excess. For instance, in the gluino production case, the best fit region for the ATLAS8 $Z$+MET excess is excluded by the ATLAS13 result. This constraint is quite robust and almost independent of the details of the mass spectrum. The CMS13 ATLAS-like channel also provides a similar and slightly stronger constraint. This CMS13 constraint corresponds to $(\epsilon\sigma)^{\rm ATLAS}_{13\text{TeV}} \lesssim 4$-$5$ fb in terms of the ATLAS13 event selection.

On the other hand, in the squark case, these constraints are not so severe. This is because the squark production cross section is relatively smaller than the gluino, and lighter squarks are favoured to explain the ATLAS8 $Z$+MET excess. In such a case, the enhancement of the cross section at the 13 TeV LHC is not so large, and these constraints are relatively weak. If the gluino mass is light enough to contribute to the squark production, a heavier squark mass is favoured. In this case, however, the dominant production comes from the valence quarks and the enhancement at the 13 TeV LHC is not so high. Therefore the $Z$+MET search by ATLAS13 and CMS13 ATLAS-like channel can be consistent with the ATLAS8 $Z$+MET excess in the squark case.

This is also the case for the heavy gluon. Since the heavy gluon has spin 1, the dominant production channel is the valence and sea quark fusion.
Then the constraints of the ATLAS13 and CMS13 results get relatively weak.

\subsubsection*{$Z$+MET by CMS8 and CMS13 SRA-B}

Generally, compatibility between the ATLAS8 $Z$+MET excess and the CMS8/13 constraints is hard. Regarding the CMS8 constraint, it uses the different event selections from the ATLAS8 and the constraints are not parallel. This CMS8 constraint reads $(\epsilon\sigma)^{\rm ATLAS}_{8\text{TeV}} \lesssim 0.3$-$0.8$ fb. This constraint excludes the $1\sigma$ best fit region in many models.

The constraints by the CMS13 SRA-B channels are also powerful in excluding the models. This constraint corresponds to  $(\epsilon\sigma)^{\rm ATLAS}_{13\text{TeV}} \lesssim 1$-$5$ fb. However, the event selection is quite different from the ATLAS8/13 searches. Especially the $b$-jet multiplicity is important. For the $Z$+MET channel, one of the main SM background comes from $t\bar{t}$ events. In fact, in the SRB region, which requires $\geq 4$ jets and provides the best sensitivity for most of the models, the observed number of the events with the $b$-jets is about twice of that without the $b$-jets. Therefore the new physics models with $b$-jets suffer from relatively weaker constraints by the CMS13 SRA-B. The heavy gluon case is such an example as seen in Fig.~\ref{fig:gheavy}. Therefore, this constraint has larger model-dependence. For instance, allowing a small nonzero branching fraction into $b$~quarks may relax this constraint. In the case of the GGM model with the gluino pair production, we assume the gluino decay into $u,d,s$, and $c$ quarks. In fact, by adding the $b$ quark channel, we find the constraints get weaker. However, in such a case, other multi-$b$ jet searches will also constrain the models and thus the situation is complicated. Detailed analysis on this is out of the scope of the present simplified model.

\begin{figure}[tbp]
\centering
\includegraphics[width=0.7\textwidth]{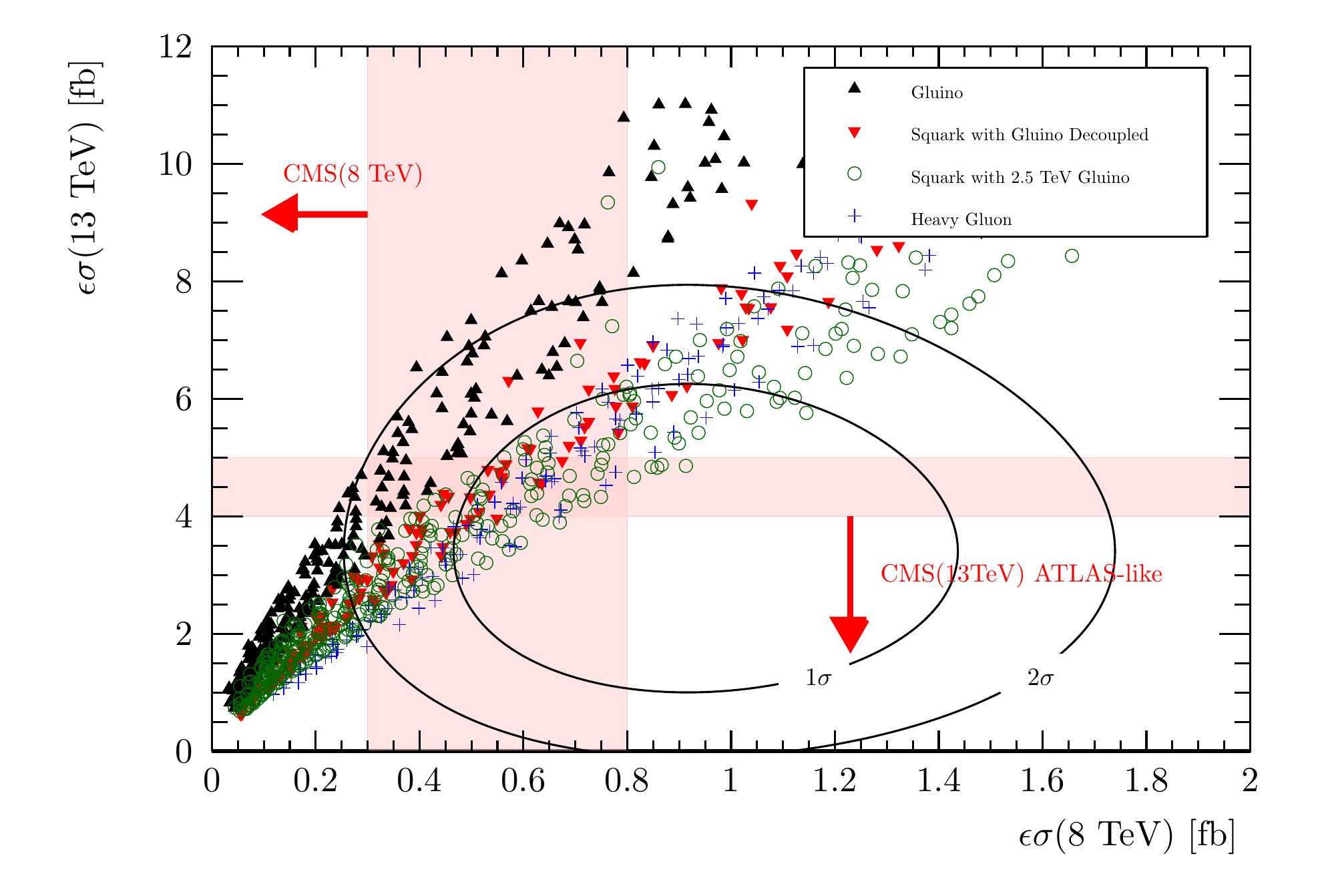}
\caption{Scatter plot of the visible cross sections $\epsilon \sigma(8~{\rm TeV})$ and $\epsilon \sigma(13~{\rm TeV})$.
 The values of $M_{\rm parent}$ and $M_{\rm NLSP}$ $(\tan \theta_3)$ are sampled with $40~{\rm GeV} \times 40~{\rm GeV}$ $(0.02)$ grid spacing in each simplified model.
 }\label{fig:scatter}
\end{figure}

In Fig.~\ref{fig:scatter}, we show a scatter plot of $\epsilon \sigma(8~{\rm TeV})$ vs $\epsilon \sigma(13~{\rm TeV})$ with favoured regions in light of the ATLAS8/13 excesses. Here we pick up the model points of $M_{\rm parent}$ and $M_{\rm NLSP}$ $(\tan \theta_3)$ with  $40~{\rm GeV} \times 40~{\rm GeV}$ $(0.02)$ grid spacing. For reference, we show the constraints from CMS8 and CMS13 (ATLAS-like), and their model dependence is shown in red shaded regions. In the gluino and squark cases, we adopt the degenerate mass spectrum and the gluino two-body decay. We see that the new physics models capable of explaining the ATLAS8 $Z$+MET excess tend to predict larger cross sections at the 13 TeV LHC than observed. In particular, the gluino case predicts too large cross sections.

\section{Discussion}\label{sec:discussion}

Since the ATLAS collaboration reported the excess in the $Z$+MET search in the 8 TeV run, many models of new physics has been proposed to explain it. In this paper, we discussed the consistency of these models with the 8 TeV and 13 TeV LHC results. We found that the 13 TeV LHC results generically give rise to the tension with the ATLAS8 $Z$+MET excess.

Among new physics models, the squark case seems to provide relatively better fitting. In this paper, we assume all the 1st/2nd-generations squarks have a common mass. If we remove this condition and instead assume that only right-handed squarks are light, the compatibility of the ATLAS8 $Z$+MET excess fitting and other constraints will be improved, since a lower mass of the squarks is preferred by the ATLAS8 $Z$+MET excess and thus the $R$ value gets smaller. Moreover, if we include the sbottom channel, the constraints of the 13 TeV LHC, particularly CMS13 SRA-B channels can be relaxed.

Although the squark cases provide relatively good fits, the constraints of ATLAS8/13 and CMS8/13 are not compatible completely. If the ATLAS8/13 excesses are true, the CMS8/13 searches would naturally have shown some excesses. This incompatibility problem is most severe in between the ATLAS13 and CMS13 (ATLAS-like) searches, since these two essentially have the same event selection cuts. In fact, the observed numbers of the total events for ATLAS13 and CMS13 seem to be consistent with each other. However, the background estimations look different. Therefore it would be possible that the incompatibility of the ATLAS8/13 and CMS8/13 $Z$+MET searches can come from the systematic difference of the SM background estimations. We hope the ATLAS and CMS collaborations investigate the compatibility of their background estimations and reveal the origin of the deviation.

Let us discuss the limitations of our simplified model descriptions. In the present analysis, we assume that the branching fraction of the parent particle into a $Z$ boson is 100\%. If this branching fraction is not 100\%, the tensions between the 8 TeV and 13 TeV $Z$+MET signals are slightly relaxed, since we need a smaller mass of the parent particle to explain the 8 TeV ATLAS $Z$+MET excess and thus the $R$ value gets smaller. However, this effect is quantitatively not so significant. For instance, when the branching fraction into a $Z$ boson is 50\%, the $R$ value gets reduced by only around 10\%. This can be inferred from Fig.~\ref{fig:scatter} because variation in the parent particle mass does not lead to significant change of the $R$ value (inclination of the scattering points). Namely, the $R$ value is dominantly determined by the production process of the parent particle. Moreover, in the case that the branching fraction into a $Z$ boson is not 100\%, the jets+MET+zero-lepton constraints get severer and one-lepton+MET or $b$-jets+MET constraints will also be stronger, depending on the other opening decay modes. Therefore we expect the present simplified descriptions to be a conservative and robust way to discuss the consistency between the 8/13 TeV LHC results in the light of the ATLAS $Z$+MET excess.

Finally we comment on the uncertainty of our signal estimations. For the consistency between the ATLAS $Z$+MET and jets+MET+zero-lepton signals, we mainly focus on the compressed mass spectra. In such cases, however, the signal estimation suffers from larger uncertainties of the event generator and detector simulations although our simulation setup is calibrated so that we can reproduce the new physics constraints provided by the ATLAS and CMS papers. For instance, using a different subversion of Pythia with the default parameters, we found up to 50\% difference in the acceptance rates for the same mass spectrum. These uncertainties give large impacts on the exclusion limits from the jets+MET+zero-lepton searches. For more accurate signal estimations, we need more careful treatment on these uncertainties. However, we expect these uncertainties give a less significant impact on the estimation of the $Z$+MET signals and our conclusion will be essentially unchanged.

\section*{Acknowledgments}

XL is supported by DOE grant DE-SC-000999. TT acknowledges JSPS for a Grant-in-Aid for JSPS Fellows and the Grant-in-Aid for Scientific Research No.~26$\cdot$10619. Parts of TT's work were done during his visit to DESY, and he is grateful for hospitality of DESY people.
\bibliographystyle{aps}
\bibliography{ref}

\end{document}